\documentclass[journal]{IEEEtran}
\usepackage[utf8]{inputenc}
\usepackage{amsmath, amssymb,amsthm, bm, algorithm2e}
\usepackage[style=base, tablename=Table]{caption}
\usepackage{array,multirow,graphicx}
\usepackage{xcolor}
\usepackage{sharedcolors}
\usepackage{mathtools}
\usepackage{wrapfig}
\usepackage[quotient-mode=fraction,binary-units]{siunitx}
\usepackage[stretch=10,shrink=10,kerning=true,tracking=true]{microtype}
\usepackage[acronym,nomain]{glossaries}
\usepackage{tikz}	
\usepackage{pgfplots}	
\usepackage{enumitem}
\usepackage{cleveref}
\usepackage{xfrac}

\DeclareSIUnit{\sample}{S}

\newcommand\spacedstyle[1]{\SetTracking{encoding=*}{0}\lsstyle}

\DeclareUnicodeCharacter{2212}{−}
\usetikzlibrary{patterns}
\usetikzlibrary{shapes.arrows, fit, positioning, calc, decorations}
\usetikzlibrary{external}
\usetikzlibrary[pgfplots.colormaps]
\pgfplotsset{compat=newest,}
\usepgfplotslibrary{groupplots}
\usepgfplotslibrary{dateplot}

\newcommand{\trans}{{\rm T}}   
\newcommand{\herm}{{\rm H}}    
\newcommand{\ma}[1]{\bm{#1}} 
\newcommand{\ten}[1]{\bm{\mathcal{#1}}} 

\newcommand{\R}{\mathbb{R}}
\newcommand{\C}{\mathbb{C}}

\newcommand{\N}{\mathbb{N}}
\newcommand{\forward}[2]{\bm{\phi}_{#1}(#2)}
\newcommand{\backward}[2]{\bm{\beta}_{#1}(#2)}

\DeclareMathOperator*{\Diag}{diag}
\DeclareMathOperator*{\BlkDiag}{blkdiag}

\DeclareMathOperator*{\Min}{min}
\DeclareMathOperator*{\Vectorize}{vec}
\DeclareMathOperator*{\Argmax}{argmax}
\newcommand{\Text}[1]{{\hspace{3mm} \text{#1} \hspace{3mm}}}
\newtheorem{theorem}{Theorem}[section]
\theoremstyle{definition}
\newtheorem{remark}{Remark}[section]
\newtheoremstyle{mystyle}{}{}{}{}{\bf}{}{\newline}{}
\theoremstyle{mystyle}
\newtheorem{discussion}{Discussion}
\newcommand{\Abs}[1]{{\left| #1 \right|}}
\newcommand{\Norm}[1]{{\left\Vert #1\right\Vert}}

\newacronym{adc}{ADC}{Analogue-to-Digital Converter}
\newacronym{arpack}{ARPACK}{ARnoldi PACKage}
\newacronym{asic}{ASIC}{Application Specific Integrated Circuit}
\newacronym{awgn}{AWGN}{Addtive White Gaussian Noise}
\newacronym{cpu}{CPU}{Central Processing Unit}
\newacronym{crb}{CRB}{Cramér-Rao-Bound}
\newacronym{cs}{CS}{Compressed Sensing}
\newacronym{dft}{DFT}{Discrete Fourier Transform}
\newacronym{fim}{FIM}{Fisher Information Matrix}
\newacronym{fmc}{FMC}{Full Matrix Capture}
\newacronym{gpu}{GPU}{Graphical Processing Unit}
\newacronym{pwc}{PWC}{Plane Wave Compounding}
\newacronym{pwi}{PWI}{Plane Wave Imaging}
\newacronym{sa}{SA}{Synthetic Aperture}
\newacronym{spw}{SPW}{Single Plane Wave}
\newacronym{ula}{ULA}{Uniform Linear Array}
\newacronym{rc}{RC}{Raised Cosine}
\newacronym{cmos}{CMOS}{Complementary Metal Oxide Semiconductor}
\newacronym{cots}{COTS}{Commercial-Off-The-Shelf}
\newacronym{das}{DAS}{Delay-and-Sum}
\newacronym[longplural={Fast Fourier Transforms}]{fft}{FFT}{Fast Fourier Transform}
\newacronym{mpu}{MPU}{Microprocessor Unit}
\newacronym{fpga}{FPGA}{Field Programmable Gate Array}
\newacronym{dsp}{DSP}{Digital Signal Processor}
\newacronym{lo}{LO}{Local Oscillator}
\newacronym{lam}{LAM}{Large Area Monitoring}
\newacronym{ndt}{NDT}{Nondestructive Testing}
\newacronym{nde}{NDE}{Nondestructive Evaluation}
\newacronym{ric}{RIC}{Restricted Isometry Constant}
\newacronym{rip}{RIP}{Restricted Isometry Property}
\newacronym{roi}{ROI}{Region of Interest}
\newacronym{saft}{SAFT}{Synthetic Aperture Focusing Technique}
\newacronym{shm}{SHM}{Structural Health Monitoring}
\newacronym{ssr}{SSR}{Sparse Signal Recovery}
\newacronym{snr}{SNR}{Signal-to-Noise Ratio}
\newacronym{ista}{ISTA}{Iterative Shrinkage-Thresholding Algorithm}
\newacronym{fista}{FISTA}{Fast Iterative Shrinkage-Thresholding Algorithm}
\newacronym{twista}{TWISTA}{Two-step Iterative Shrinkage-Thresholding Algorithm}
\newacronym{stela}{STELA}{Soft-Thresholding with Exact Line Search Algorithm}
\newacronym{rd}{RD}{Random Demodulator}

\newtheorem{definition}{Definition}
\begin{document}
\title{Frequency Sub-Sampling of Ultrasound Non-Destructive Measurements: Acquisition, Reconstruction and Performance}

\author{Jan~Kirchhof,~\IEEEmembership{Student Member,~IEEE,}
        Sebastian~Semper,
        Christoph~W.~Wagner,
        Eduardo~Pérez,~\IEEEmembership{Student Member,~IEEE,}
        Florian~Römer,~\IEEEmembership{Senior Member,~IEEE,}
        and~Giovanni~Del~Galdo,~\IEEEmembership{Member,~IEEE}
        \thanks{%
                Sebastian~Semper is funded by DFG under the project ``HoPaDyn'' Grant-No. TH 494/30-1.
                Christoph~W.~Wagner is funded by Carl-Zeiss-Stiftung under the project ``PRIME''. Eduardo Pérez is funded by DFG under the project ``CoSMaDU'' with grant GA 2062/5-1.
                This work was also supported by the Fraunhofer Internal Programs under Grant No. Attract 025-601128.
                Manuscript received \today .
        }
}
\markboth{tbd}%
{tbd}


\maketitle

\begin{abstract}
    In ultrasound~\gls{ndt}, a widespread approach is to take synthetic aperture measurements from the surface of a specimen to detect and locate defects within it. Based on these measurements, imaging is usually performed using the \gls{saft}.  However, \gls{saft} is sub-optimal in terms of resolution and requires oversampling in time domain to obtain a fine grid for the \gls{das}. On the other hand, parametric reconstruction algorithms give better resolution, but their usage for imaging becomes computationally expensive due to the size of the parameter space and the large amount of measurement data in realistic 3-D scenarios when using oversampling. \par
    In the literature, the remedies to this are twofold: First, the amount of measurement data can be reduced using state of the art sub-Nyquist sampling approaches to measure Fourier coefficients instead of time domain samples. Second, parametric reconstruction algorithms mostly rely on matrix-vector operations that can be implemented efficiently by exploiting the underlying structure of the model. \par
    In this paper, we propose and compare different strategies to choose the Fourier coefficients to be measured. Their asymptotic performance is compared by numerically evaluating the \gls{crb} for the localizability of the defect coordinates. These subsampling strategies are then combined with an $\ell_1$-minimization scheme to compute 3-D reconstructions from the low-rate measurements. Compared to conventional \gls{das}, this allows us to formulate a fully physically motivated forward model matrix. To enable this, the projection operations of the forward model matrix are implemented matrix-free by exploiting the underlying 2-level Toeplitz structure. Finally, we show that high resolution reconstructions from as low as a single Fourier coefficient per A-scan are possible based on simulated data as well as on measurements from a steel specimen.
\end{abstract}
\glsresetall
\section{Introduction \label{sec:intro}}

\subsection{State of the art}
In ultrasound~\gls{ndt} defects are detected and localized by inserting an ultrasonic pulse into a specimen and collecting the resulting echo signals~\cite{krautkramer_springer_2013}. A typical measurement setup consists of a single transducer or transducer array that is used both as transmitter and receiver. In the single channel case the transducer is moved on the specimen surface and measurements are collected at each scanning position to form a so-called synthetic aperture. \par
The measurements in these setups are usually acquired in time domain and sampled at a frequency much higher than the Nyquist frequency to create a dense time grid that the subsequent \gls{das} can use to achieve a well-resolved reconstruction. An image of the specimen interior is then commonly computed based on the \gls{saft}~\cite{spies_jnde_2012} or its multi-channel extension~\cite{holmes_ndtint_2005}. More recently, model based approaches have been introduced, where the main idea is to treat the reconstruction as an inverse problem based on a physically derived model matrix that is solved using standard tools from linear algebra~\cite{berthon_physicsinmedicine_2018}. This enables us to include more complex mechanisms of wave propagation that cannot be captured by simple \gls{das} such as attenuation and temporal dispersion~\cite{carcreff_ieeeuffc_2014}, or acoustic shadowing~\cite{kirchhof_ieeeicassp_2017}. Additionally one can account for the pulse shape or elementary signature~\cite{quaegebeur_ultrasonics_2012}, which is usually modeled as the real part of a Gaussian windowed sinusoid~\cite{demirli_ieeeuffc_2001}. \par
Oversampling in time domain leads to a large amount of measurement data, which results in a computationally expensive reconstruction, especially when it comes to 3-D imaging~\cite{kirchhof_ieeeius_2018}. However, parametric models enable the use of \gls{cs} techniques~\cite{candes_ieeeit_2006} to reduce the amount of measurement data. Initially, mostly random Gaussian measurement kernels have been investigated~\cite{quinsac_advancesacousticsvibration_2012}, since for those, theoretical guarantees could be established ensuring that the reconstruction is robust and stable. This entails that mostly theoretical results exist in the ultrasound literature, as the design of a generic \gls{cs} hardware is a challenging task~\cite{mangia_ieeecircuits_2020}. \par
To reduce the amount of measurement data per incoming echo signal, prior work has shown that highly structured signals such as ultrasonic recordings can be completely recovered in frequency domain from samples taken at rates significantly below Nyquist, provided that they are measured with an appropriate sampling architecture~\cite{eldar_cambridge_2015}.
Based on the recovered Fourier coefficients, different reconstruction strategies can be employed. One approach is to use beamforming in frequency domain~\cite{chernyakova_ieeeuffc_2014}. Another approach is to formulate the reconstruction as an inverse problem based on a physically motivated forward model and extend this model by the compression scheme. In the \gls{cs} context, sub-sampled Fourier measurements can guarantee stable and robust reconstruction since a randomly generated partial Fourier matrix fulfills the D-\gls{rip}~\cite{candes_DRIP_2010,krahmer2011ward,haviv2015fourierrip}. It has been shown that high quality 3-D images based on synthetic aperture measurements can be reconstructed from only a small number of Fourier samples collected at each measurement position~\cite{semper_eusipco_2019}.
Additionally, the acquisition of Fourier coefficients is an attractive choice, as hardware architectures to measure them directly already exist~\cite{lehne_phd_2008,tur_ieeetsp_2011,spaulding_ieeeicassp_2015,mulleti_ieeetsp_2020}. In multi-channel setups, the compression is often realized by sub-sampling the channels of a large array~\cite{besson_ieeeicip_2016,ramkumar_ieeeuffc_2020,perez_ieeeius_2019}, which exploits the same subsampling principles as they are used in sparse array design. \par
Generally, compressed observations are post-processed by means of \gls{ssr} to solve the parameter estimation problem at hand. Often, the sparsity is assumed in the parameter domain as the number of defects inside the specimen is small~\cite{Tuysuzoglu2012sparsity_ultrasound}. The actual solution is usually computed using greedy~\cite{pati_asilomarcon_1993} or $\ell_1$-minimization~\cite{beck_siam_2009,yang_ieeetsp_2018} algorithms. These techniques have proven to produce images with superior quality~\cite{quaegebeur_ultrasonics_2012} and better resolution of closely spaced defects~\cite{laroche_ieeeuffc_2020} compared to images using only \gls{das}.\par
To reduce the computational effort that arises when aiming for high estimation quality, efficient implementations exploiting the structure of the involved linear mappings need to be employed~\cite{karimi_jasa_2017}. Often these structure exploiting algorithms allow a high degree of parallelization so that further speed-ups are possible by the use of a \gls{gpu}~\cite{kirchhof_ieeeius_2018, bueno_journalofrealtimeprocessing_2018}. \par
Additionally, a parametric model enables to analyze the influence of the parameters on the measurements. One way to do so is to look at the estimation variance by means of the \gls{crb}~\cite{kay_prenticehall_1993}. In the ultrasound field, the \gls{crb} has been used to quantify time delay estimation jitter and the impact of the choice of bandwidth and center frequency of ultrasound pulse-echo measurements~\cite{walker_ieeeuffc_1995, viola_ieeeuffc_2003}, and the achievable resolution when locating single point-scatterers~\cite{desailly_physicsinmedicine_2015}. Moreover, the \gls{crb} has been used as a criterion for array design and sensor placement~\cite{an_ieeeicassp_2020, gershon_ieeeicassp_2020} as well as to optimize spatial sub-sampling patterns in \gls{cs}~\cite{perez_ieeeicassp_2020}.
\subsection{Motivation}\label{sec:motivation}
The main focus of this paper lies on synthetic aperture ultrasound measurements in~\gls{ndt}. A single transducer is placed on the planar surface of a specimen and moved to different positions on a regular 2-D grid. Specifically, we are interested in scenarios where a high resolution 3-D reconstruction based on these measurements is required, for example to resolve closely spaced scatterers. Since the goal is to detect defects, we need to scan the whole region of interest leading to a large number of measurement positions and therefore a large amount of measurement data. \par
Inspection measurements can be described using a three-stage model, comprised of (1) data management (collection and storage of measurement data), (2) sense making (data analysis, analytics and feature extraction) and (3) decision making~\cite{cawley_structural_health_monitoring}.
From stage to stage, data volume reduces while data value increases. \par
For manual inspections, a trained technician usually performs tasks of all model stages in full during inspection.
However, expertise is only required for decision making in the final stage, leaving the (on-site) data-management and (off-site) sense-making stages up for automation.
As a result, the required effort for on-site inspections can be reduced greatly.
However, since expertise is no longer involved prior to the final decision making stage, data filtering also can no longer be applied in earlier stages.
This not only leaves an overwhelming amount of unfiltered data to handle, but also clogs the computationally intense processes of the sense-making stage.
To illustrate the amount of data gathering required, a small example size with a scanning grid of $100 \times 100$ locations and an A-scan time-length of $1000$ samples is already sufficient.
With~\SI{16}{\bit} data quantization, one full synthetic aperture measurement requires~\SI{20}{\mega\byte} of data.
Further assuming $100$ different locations for a single inspection, a total volume of~\SI{2}{\giga\byte} accrues.
To reduce the amount of measurement data, a natural approach in this scenario is to minimize the amount of measurement samples per scan position without losing relevant information.
Instead of measuring on a dense grid in time domain (sampling frequencies above~\SI{100}{MHz} are common, although the ultrasound pulse rarely exceeds a bandwidth of, say,~\SI{10}{MHz}), we can use existing hardware architectures~\cite{tur_ieeetsp_2011,mulleti_ieeetsp_2020} to obtain Fourier coefficients from samples taken at a much lower rate. \par
The adoption of CS enables the modification of the three-stage model on the hardware side so that it mimics the tasks of the technician. Complex post-processing routines are only necessary as a final step prior to decision making, meaning the operations performed by the sensors can be streamlined. Reducing the data rate directly at the measurement stage has the added benefits of diminishing power requirements and enabling data streaming to a possibly remote processing unit~\cite{mamistvalov2020sparse}. Further, progression toward the so-called NDE4.0~\cite{valeske2020next} increases the attractiveness of embedded sensors for \gls{shm}~\cite{xinlin_builtin_shm_sensor_network} scenario in which power consumption severely constrains active sensors. \par
A naturally arising question is: how many and which measurements are necessary for a given sub-sampling methodology to result in robust and stable reconstruction?
Considering the pulse-echo of a single volume-element within the specimen, an answer to this question is already indicated by the observation that the echo caused by said volume-element is systematically represented in a large number of adjacent A-scans.
Describing the linear measurement model in all three spatial dimensions reveals a structure presenting large amounts of spatial redundancy between A-scans.
In~\cite{semper_eusipco_2019}, the model is extended by sub-selecting only a few Fourier coefficients from each A-Scan, effectively representing a compression of the measurement data. The parameter estimation problem of finding defect positions based on these compressed measurements is formulated as an \gls{ssr} problem.
As this exhibits a high dimensional parameter space, it quickly becomes computationally expensive or even intractable.
Standard linear algebra solvers that rely on explicit representations of the system matrix, already break down on moderate problem sizes, since dense or even sparse representations of the operator matrix quickly become infeasibly large. As illustration, for the given example the operator matrix is of size $(1000 \cdot 100^2)^2$, even exceeding the memory capabilities of current mainframe systems.
Instead, a matrix-free representation of the operator is possible that exploits the embedded structure of the matrix in a computationally- and memory-efficient way.
This implementation is designed to be flexible in terms of modeling parameters (e.g. pulse shape, dimensions of the specimen, grid dimensions of the \gls{roi}) as well as the measurement strategy (different compression strategies, and/or uncompressed measurements). \par
In addition, most of the existing algorithms which rely only on matrix-vector products usually have side constraints that implicitly require more knowledge about the matrix. Therefore, implementations need to be tweaked to account for this. To give an example, the \gls{fista} requires the largest singular value of the system matrix in order to select the correct step size, which cannot be straightforwardly computed without the full matrix available. Although an approximation using \gls{arpack} is possible in our framework, this adds a large computational overhead that in some cases should be omitted. Lastly, the implementation should be easily usable with different computation setups available, e.g. likewise on a \gls{gpu} or \gls{cpu}. \par
In order to precisely quantify the effects of our estimation procedure, we investigate the proposed methodology based on the asymptotic and real-world performance. The asymptotic performance is given in terms of the \gls{crb} of the resolution of locating single point-scatterers. This is helpful, as the \gls{crb} directly quantifies how parametrizing the measurement setup, e.g. how many Fourier coefficients are measured, influences the parameter estimation. The real-world performance is evaluated based on the reconstruction of simulated as well as real ultrasound measurement data.
\subsection{Main contributions}
In this paper, we develop and investigate a complete framework for high resolution ultrasound imaging based on a small number of Fourier measurements and a complete 3-D propagation model. 
To obtain these measurements, we propose novel sampling strategies and compare their performance to existing approaches including our previous work~\cite{semper_eusipco_2019}. \par
We first derive a theoretic scaling law for synthetic aperture ultrasound measurements using the strategy of sampling \emph{uniformly at random} given by the \gls{cs}~literature~\cite{haviv2015fourierrip}. Specifically, we show that the required number of Fourier coefficients in this case only depends on the worst case sparsity of all scans, i.e. the scanning position that sees the maximum number of echoes, showing that it does not exploit the correlation between adjacent scans.   
In comparison, the novelty of the sampling strategies proposed in our work is twofold. First, we propose to incorporate prior knowledge about the spectrum of the inserted pulse into the design of the sampling pattern. This sparks two possible strategies, which we term \emph{maximal} and \emph{random energy-based} sampling. They alleviate the need for additional random sign-flips prior to the Fourier sampling as dictated by \gls{cs} theory~\cite{haviv2015fourierrip} and therefore simplify the hardware requirements when implementing the Fourier subsampling in the analog domain. Second, we make full use of the complete 3-D model by varying the sampling pattern at each scan position using the \emph{random energy-based} strategy. 
The strong spatial correlation between the measurements at adjacent scan positions leads to a trade-off between temporal and spatial measurements.
However, since we need to scan the specimen at a certain minimum density to ensure defect detection, this can mainly be exploited to reduce the number of temporal samples. 
In fact, we show numerically that when the number of spatial scanning positions is large, only taking a single (but varying) Fourier coefficient per scan position does not substantially decrease the \gls{crb} compared to the uncompressed case in a single defect scenario (or a scenario where the distance between defects is sufficiently large). \par
The ultrasound measurements are modeled using a parametric forward model that conveys both the ultrasound propagation as well as the received pulse shape. The analytic signal is used instead of only modeling the real-valued RF signal. \par
The matrix-free implementation for the reconstruction is achieved by exploiting the block-wise 2-level Toeplitz structure of the forward model.
This enables to implement the effect of both the subsampling and the model in the reconstruction using only \glspl{fft} and indexing operations.
The implementation is done in Python using the fastmat package~\cite{wagner_arxiv_2017}.
The reconstructions are carried out via the \gls{fista} algorithm.
The resulting matrix-free operator can not only be used for high resolution parametric reconstruction but also, by applying its adjoint on the measurements, yields a single-step ``compressed \gls{saft}`` reconstruction that considers our assumptions about the model and the compression scheme.\par
With this implementation at hand, using simulated as well as measurement data it is shown  that the spatially randomized strategies allow to produce 3-D images from a single Fourier coefficient per A-scan allowing precise localization and sizing of several test defects.\par
The remainder of the paper is organized as follows:
In Sec.~\ref{sec:data_model} we derive the ultrasound propagation model used throughout this paper. 
In Sec.~\ref{sec:fourier_subsampling}, we introduce the novel Fourier acquisition schemes and derive a theoretic scaling law based on \gls{cs} theory. 
In Sec.~\ref{sec:algs}, we discuss the reconstruction process and provide concrete matrix-free algorithms for the implementation of the model and compression operator. 
Using \gls{fista} as our example algorithm, we discuss practical solutions to approximate the required largest singular value based on a matrix-free software implementation. 
In Sec.~\ref{sec:implementation}, we compare the hardware requirements as well as the computation complexity of the proposed \gls{cs} architecture compared to state of the art systems. 
In Sec.~\ref{sec:asymptotic_performance}, we derive the single-scatterer \gls{crb} for the model and investigate the influence of the compression on the localization capability of the measurements asymptotically. 
Then, in Sec.~\ref{sec:num_sims}, we provide example reconstructions from numerical simulations as well as using realistic measurement data to back up the theoretic claims. 
Finally, Sec.~\ref{sec:Conclusion} concludes the paper. 

\section{Observation Model \label{sec:data_model}}
\subsection{Ultrasound data model \label{sec:us_data_model}}
We consider a pulse-echo setup, where a single transducer is used to insert an ultrasonic pulse $h : \R \rightarrow \R$ with $t \mapsto h(t)$ into the specimen, which we also assume to be the received pulse. This means we assume dispersion in the medium to be negligible. The specimen is considered to be homogeneous and isotropic with constant speed of sound $c \in \R^+$ and to possess a flat surface. It contains $D \in \N$ point-like defects located at unknown positions $(x_d, y_d, z_d) \in \R^3$ for $1 \leqslant d \leqslant D$ that are to be localized. We assume that we can omit reflections from known features of the specimen (such as the back wall) by appropriate windowing, such that only the $D$ reflections from the defects remain.

We first introduce a continuous model for the observations. If we define $f^{(a)}$ as the analytic signal of a function $f$ via
\begin{equation}
    f^{(a)}(t) = f(t) + \jmath \mathcal{H}\{f(t)\},
\end{equation}
where $\mathcal{H}\{\cdot\}$ is the Hilbert transform of $f$, the analytic noiseless signal $b^{(a)}_{x,y} : \R \rightarrow \C$ received by the transducer from position $(x,y) \in \R^2$ can be modeled as
\begin{align}\label{eq:a_scanmodel}
    b^{(a)}_{x,y}(t) &= \sum_{d=1}^D a_d  g_{x,y}(x_d, y_d, z_d)\cdot \nonumber \\
    &\phantom{{}=\text{Re}\lbrace} \vphantom{\sum_{d=1}^D} h^{(a)}(t - \tau_{x,y}(x_d, y_d, z_d)).
\end{align}
Here, $a_d \in \C$ is the complex reflectivity and $\tau_{x,y} : \R^3 \rightarrow \R^+$ is the time of flight from the transducer at sample position $(x,y) \in \R^2$ to the $d$-th reflector and back. It can be computed as
\begin{equation}
    \tau_{x,y}(x_d, y_d, z_d) = \frac{2}{c}\sqrt{(x - x_d)^2 + (y -y_d)^2 + z_d^2}.
\end{equation}
Additionally, $g_{x,y}:\R^3 \rightarrow \R^+$ represents the transducer characteristic, which models its directivity towards $(x_d, y_d, z_d)$. In frequency domain, we obtain
\begin{align}\label{eq:a_scanmodel_freq}
    B_{x,y}(\omega) &= \sum_{d=1}^D a_d g_{x,y}(x_d, y_d, z_d) H(\omega) \cdot {\rm e}^{-\jmath \omega\tau_{x,y}(x_d, y_d, z_d))},
\end{align}
where the function $H : \R \rightarrow \C$ is the Fourier transform of the pulse function $h$. To simplify the notation we introduce the atomic functions
$f_{x,y}(x_d, y_d, z_d, t) : \R^4 \rightarrow \C$ as
\begin{equation}
f_{x,y}(x_d, y_d, z_d, t)
  = g_{x,y}(x_d, y_d, z_d) h^{(a)}(t - \tau_{x,y}(x_d, y_d, z_d))
\end{equation}
and we can now write concisely
\begin{align}\label{eq:a_scanmodel_short}
    b^{(a)}_{x,y}(t) &= \sum_{d=1}^D a_d \cdot f_{x,y}(x_d, y_d, z_d, t).
\end{align}
Next, we transform the continuous model into a discrete one. This consists of several steps. First, we naturally have to assume a discrete and finite set of observation locations. Second, we discretize the received signals by means of Nyquist rate sampling. Note that this does not mean we actually need to have access to this sampled observation, but instead we use it as a discrete and finite, hence convenient, representation of the continuous signal. One advantage of this representation is that it allows us to write linear transforms on the signal as matrices. Finally, we also make the same assumptions about the reconstructed signal. It is composed of defects residing on the same grid as we used for the observations. These regularity assumptions about the grid are necessary for efficient recovery to be possible as we see later on.

To define the observation grid we take synthetic aperture measurements at positions $(x, y, z=0) \in \R^3$ located on the surface of the specimen lying on an equidistant grid defined as
\begin{align*}
    G_{2D} = \lbrace (x,y) |
        x = n_x  \cdot \Delta x, n_x \in
            \left\lbrace 0, \dots, N_x-1 \right\rbrace, \\
        y = n_y \cdot \Delta y, n_y \in
            \left\lbrace 0, \dots, N_y-1 \right\rbrace \rbrace ,
\end{align*}
where $\Delta x = \Delta y$ is the grid spacing and $N_x \in \N$ and $N_y \in \N$ are the number of samples in each spatial dimension.

After discretizing the observations $b^{(a)}_{x,y} : \R \rightarrow \C$ with a sampling rate $t_s = \frac{1}{f_s}$ to vectors $\bm b_{x,y} \in \C^{N_t}$, resulting in $N_t$ samples, the full set of $N_x \times N_y$ measurements can be combined into a 3-D array as
\begin{equation}
    \ma{B} = \left[\begin{array}{cccc}
      \ma{b}_{0,0} & \ma{b}_{0,1} & \dots & \ma{b}_{0, N_y-1} \\
      \vdots & & \ddots & \vdots \\
      \ma{b}_{N_x -1 ,0} & \ma{b}_{N_x - 1,1} & \dots & \ma{b}_{N_x-1, N_y-1} \\

      \end{array}\right],
\end{equation}
such that $\bm B \in \C^{N_x \times N_y \times N_t}$ by stacking the vector $\bm b_{i,j}$ into the third indexing dimension. To define the considered defects' locations we extend $G_{2D}$ along the remaining spatial $z$ dimension and define the 3D grid
\begin{equation*}
    \begin{split}
        G_{3D} = \lbrace (x, y, z) \vert &(x,y) \in  G_{2D}, \\
        &z = n_z \cdot \Delta z, n_z = 0, \dots, N_t-1 \rbrace
    \end{split}
\end{equation*}
with $\Delta z = t_s \cdot c$. This aligns the defect locations' $z$-coordinates with the time sampling of the observations resulting in a total number of $N = N_x N_y N_t$ observations and consequently also $N = N_x N_y N_z$ possible defect positions.

It is worth noting that the gridding process introduces an inherent modeling error, which is negligible as long as we choose $G_{3D}$ such that the Nyquist rate is obeyed in all spatial dimensions.

In order to simplify the observation space to being one dimensional we define the vectorized version of $\bm B$ as $\Vectorize{\ma{B}}$. This vectorization happens by means of
$[\Vectorize{\ma{B}}]_i = [\bm B]_{i_x, i_y, i_t}$, where $i = i_t \cdot N_x \cdot N_y + i_y \cdot N_x + i_x$ such that the index for the last dimension is varying slowest and the one for the first fastest. The same discretization process as for the observation and reconstruction locations as well as the sampling of the functions along time can be applied to the atomic functions $f_{x,y}$ and we define a matrix
\begin{equation}\label{eq:dictionary}
  \bm H_{i, j} = f_{n_{x,i}\Delta x, n_{y,i} \Delta y}(
    n_{x,j} \Delta x, n_{y,j} \Delta y, n_{z,j}\Delta z, n_{t,i} t_s
  ),
\end{equation}
where $i = n_{x,i} N_y N_z + n_{y,i} N_z + n_{t,i}$ and $j = n_{x,j} N_y N_z + n_{y,j} N_z + n_{z,j}$ realize the same vectorization of $f_{x,y}$ as with $b_{x,y}$. In other words, the column $\ma{H}_{\cdot,j}$ contains the vectorized and discretized volumetric observation of a single reflector at (vectorized) position $j$. This is expressed in discrete time domain concisely via
\begin{equation}
    \ma{b} = \Vectorize\{\ma{B}\} = \ma{H} \ma{a}
\end{equation}
and as such it is the basis for the following introduction of the sampling scheme. Note that compared to \eqref{eq:a_scanmodel_short} the vector $\bm a \in \C^N$ now contains $D$ non-zero elements with value $a_d$ at unknown positions.
\subsection{Data Acquisition\label{sec:acq_model}}
The vector $\ma{b} \in \C^N$ contains the discrete time samples of A-scans from all different measurement positions stacked on top of each other. However, we wish to consider compressed observation of the A-scans. So, instead of measuring $\ma{b}$ directly, we employ a compression step and  measure a subsampled version of it. Conventionally, one assumes that one has access to values of linear functionals that are applied to the signals of interest \cite[Eq.~(1.1)]{candes_ieeeit_2006b}. One can express this as a matrix-vector product, where each row of the matrix is identified with a linear functional. The compressed measurement is given by
\begin{equation}\label{compressed_model}
    \ma{y} = \ma{\Phi} \cdot \ma{b} + \ma{n} = \ma{\Phi H a} + \ma{n}.
\end{equation}
Here, $\ma{n} \in \C^N$ represents measurement noise with currently unspecified distribution. If we consider a single A-scan we consequently have
\[
\bm y_{n_x, n_y} = \bm \Phi_{n_x, n_y} \bm b_{n_x, n_y} + \bm n_{n_x, n_y}.
\]
For the compression matrix we assume it has the structure
\[
  \bm \Phi_{n_x, n_y} = \ma{S}_{n_x, n_y} \ten{F} \ma{\Sigma}_{n_x, n_y}
\]
where $\ten{F}$ is the \gls{dft} matrix, $\ma{S}_{n_x, n_y} \in \R^{n_f \times N_t} $ is a selection matrix selecting $n_f$ out of the $N_t$ Fourier coefficients, so $\bm S$ is a row-subselected identity matrix. Moreover, $\ma{\Sigma}_{n_x, n_y}$ is a full rank diagonal matrix, a so called mixing matrix \cite{haviv2015fourierrip}, \cite{krahmer2011ward}. Summarizing, the compression consists of first mixing followed by a DFT and then finally subselecting a few Fourier coefficients. This merely depicts the mathematical model and we deal with possible hardware implementations in Sec.~\ref{sec:implementation}.

Considering the full vectorized, discretized and compressed observation, we can write
\begin{equation}\label{general_compression}
    \ma{y} = \bm\BlkDiag\{\bm \Phi_{1,1}, \dots, \bm \Phi_{N_x,N_y}\} \cdot \ma{b} + \ma{n}
\end{equation}
as a linear model for the compressed observations. Here $\bm\BlkDiag$ denotes the block diagonal operator, which aligns the matrices in its argument as blocks on the diagonal of a matrix containing zeros for all other entries.

\begin{discussion}
\begin{itemize}[wide=0.5em, leftmargin =*, nosep, before = \leavevmode\vspace{-\baselineskip}]
\item We would like to stress the fact that the following three sections on the data acquisition, algorithms, and hardware implementations are independent on some parts of the specific model presented above. It would be straightforward to use a frequency dependent transducer characteristic $g$, adopt to a slightly altered geometric setup by defining a different time of flight $\tau$ or define different grids for the sampling and reconstruction positions.
\item However, the structure of $\bm H$ we derive in Sec.~\ref{sec:algs} depends on the sampling and reconstruction positions defined via $G_{3D}$. For different acquisition schemes one would have to study the properties of $\bm H$ in order to derive efficient reconstruction algorithms.
\item We assume that the pulse shape $h$ follows a known model. Replacing it by a measured pulse shape (e.g. from a backwall echo) of a known reference measurement is another valid option. Further, the choice $h(t) = \delta(t)$ leads to $\ma{H}$ becoming a discrete \gls{das}-operator (cf. Remark~\ref{rem:saft}).
\item Still, already the ``simple'' model of this section represents realistic measurement data well, as validated in Sec.~\ref{sec:synthetic_aperture_measurements}.
\item The spectrum of any realistic $h(t)$ decays exponentially for $|f| \rightarrow \infty$. For that reason, the \gls{dft} is a valid proxy to model the Fourier coefficients, since no (practically relevant) aliasing can occur.
\item The matrix $\ma{\Sigma}_{n_x, n_y}$ is necessary for some sampling strategies. In some sense it results in the vectors $\ma{\Sigma}_{n_x, n_y} \bm b_{n_x, n_y}$ entries being centered random variables, which allows proofing reconstruction results in this setting, which are based on concentration of measure results from probability. 
\end{itemize}
\end{discussion}

\section{Fourier Subsampling \label{sec:fourier_subsampling}}
Based on the proposed data acquisition scheme, we can distinguish different subsampling strategies, which we will discuss in the following. Here, we first focus on the mathematical formulation and an analysis from the signal processing perspective. Hardware considerations are deferred to Sec.~\ref{sec:implementation}.

\subsection{Strategies}

\subsubsection{Random Sampling}

Classically, the theory of compressed sensing started out by deriving reconstruction results for random matrices~\cite{candes_ieeeit_2006}. In the proposed Fourier subsampling approach this translates to the diagonal entries of $\bm \Sigma_i$ following an independent Rademacher distribution, so $\ma{\Sigma}_i = \Diag\{\ma{\xi}_{i}\}$ for a Rademacher vector $\ma{\xi}_{i} \in \{-1, 1\}^{N_t}$. For a Rademacher vector, the entries are drawn independently from the uniform distribution on $\{-1, 1\}$. Additionally, each $\bm S_i$ selects from the uniform distribution on all subsets of $\{1, \dots, N_t\}$ with magnitude $n_f$. If we select this subsampling strategy, we indicate it with a subscript of the compression matrix via $\bm \Phi_{\rm rnd}$.

\subsubsection{Maximal Sampling}

The random sampling approach neglects any prior knowledge one has about the inserted pulse and its spectrum. One way to improve this would be to consider $\hat{\bm h} \in \C^{N_t}$ as the \gls{dft} of the inserted pulse and to define
\[
  J_q = \Argmax_{q,n} \left\{\Abs{\hat{\bm h}_n}\right\}
\]
as the index set of $\hat{\bm h}$ that refers to the $q$ entries with largest amplitude. Now we set $\bm \Sigma_i = \bm I$ for all $i \leqslant N_x \cdot N_y$ and each $\bm S_i$ such that it subselects the entries in $J_{n_f}$. This subsampling strategy is denoted as $\bm \Phi_{\rm max}$.

\subsubsection{Random Energy-based Sampling}

The choices $\bm \Phi_{\rm rnd}$ and $\bm \Phi_{\rm max}$ represent two very different approaches. The former one focuses entirely on randomness and neglects any prior knowledge about the pulse, whereas the latter focuses on the pulse only and as such drops the favorable reconstruction properties imposed by random subsampling. As a trade-off between the two strategies, we impose a random sampling based on the energy distribution of the inserted pulse. Let $q_i$ for $i \leqslant N_t$ be the normalized energy of the pulse in the $i$-th \gls{dft}-coefficient, so we have
\[
  q_i = \Abs{\hat{\bm h}_i} / \Norm{\hat{\bm h}}_1.
\]
Now, we iteratively construct a set of indices $J$ based on $\bm q \in \R^{N_t}$. Assume we have already drawn $n < n_f$ indices from the set $\{1, \dots, N_t\}$ into the set $J_n$. Then we set $\bm q_{J_n} = 0$ and then normalize $\bm q$ such that the remaining entries sum up to $1$. Then we draw another index $j_{n+1}$ and set $J_{n+1} = J_n \cup \{j_{n+1}\}$ and iterate until $n = n_f$. We again set $\bm \Sigma_i = \bm I$ and denote the compression matrix selected according to this strategy with $\bm \Phi_{\rm nrg}$.

\subsubsection{Independent A-Scans}

First, one notices that during the treatment of the individual A-scans it is possible to choose $\bm S_{n_x, n_y} = \bm S_0$ for all $n_x, n_y$. This implies that we collect the same Fourier coefficients for all A-scan positions. In case we do \emph{not} keep them fixed, we add a subscript to the compression matrix as $\bm \Phi_{\dots, f}$. The same distinction can be made for the choice of the mixing done by $\bm \Sigma_{n_x, n_y}$. If we vary them across A-scans we indicate this via $\bm \Phi_{\dots, m}$. Note that $\bm \Phi_{\dots, m,f}$ is also possible. In case of the same mixing and subsampling for each A-scan, we can also write \eqref{general_compression} more concisely
\[
  \bm \Phi = (\bm S \ten{F} \bm \Sigma) \otimes \bm I_{N_x} \otimes \bm I_{N_y},
\]
where $\otimes$ defines the Kronecker product, to describe the compression matrix. Table~\ref{tab:samp_strat} summarizes all possible combinations.

\begin{table}
  \begin{tabular}{cc|c|c}
    & &\multicolumn{2}{c}{$\bm S_{n_x, n_y} = \bm S_0$ }\\
    & &yes & no \\
    \hline
    \multirow{2}{*}{$\bm \Sigma_{n_x, n_y} = \bm \Sigma_0$} & yes &  $\bm \Phi_{\rm max}$, $\bm \Phi_{\rm nrg}$, $\bm \Phi_{\rm rnd}$ & $\bm \Phi_{{\rm nrg}, f}$, $\bm \Phi_{{\rm rnd}, f}$ \\
    \cline{2-4}
    & no & $\bm \Phi_{{\rm rnd}, m}$ & $\bm \Phi_{{\rm rnd},m,f}$
  \end{tabular}
  \caption{Sampling strategies considered in this work \label{tab:samp_strat}}
\end{table}
\subsection{Performance Guarantees for Random Subsampling \label{sec:performance_guarantee}}
Now we study the ramifications of the proposed Fourier subsampling strategies. In fact, we are able to establish some analytic results for one of the presented approaches.  In this chapter we focus on the case where $\bm \Phi = \bm \Phi_{\rm rnd}$, so each A-scan is pre-multiplied with the same $\bm \Sigma$ and we pick the same Fourier coefficients by means of $\bm S$, so that
\[
  \bm \Phi = (\bm S \ten{F} \bm \Sigma) \otimes \bm I_{N_x} \otimes \bm I_{N_y}.
\]
The recovery performance of sparsity exploiting algorithms and the used compression strategy can be measured in terms of the \gls{rip}, if a signal $\bm b$ is sparse in an orthonormal basis $\ma{H}$, which means that in $\bm b = \bm H \bm a$ the vector $\bm a$ is sparse. If on the other hand $\bm a$ is sparse, but $\bm H$ is not a basis anymore, but an overcomplete dictionary as in our case, then the so-called $\bm H$-\gls{rip}~\cite{candes_DRIP_2010} yields the natural framework for reconstruction guarantees.

If we consider a single A-scan vector $\bm b_{n_x, n_y}$ recorded at an arbitrary but fixed measurement position $(x,y)$ the pulse-echo model in \eqref{eq:a_scanmodel} results in $\bm b_{n_x, n_y}$ being a linear superposition of shifted versions of the inserted pulse. It follows that $\bm b_{n_x, n_y} = \bm G \bm \alpha_{n_x,n_y}$, where the columns of $\bm G$ are the discretized and shifted versions of the inserted pulse. In other words: the matrix $\bm G$ is the sparsifying dictionary for individual A-scans. Now, according to our compression scheme, we have $\bm y_{n_x,n_y} = \bm S \ten{F} \bm \Sigma \bm G \bm \alpha_{n_x,n_y}$ for our compressed observations.

Since the total number of defects in the specimen is assumed to be small, i.e. $\bm a$ in \eqref{eq:a_scanmodel} being sparse, we have that each $\bm \alpha_{n_x,n_y}$ is sparse as well. This implies that each $\bm b_{n_x,n_y}$ is sparse in the dictionary $\bm G$. Let now $s_{\rm max} = \max_{n_x,n_y} \Norm{\bm \alpha_{n_x,n_y}}_0$ be the maximum encountered sparsity-level over all A-scans. However, the sparsity is not prevalent with respect to a basis but to the overcomplete dictionary $\bm G$. This sparks the need for a modified reconstruction guarantee presented in the next definition.

\begin{definition}[$\bm G$-RIP,~\cite{candes_DRIP_2010}]
Let $U_k$ be the union of all subspaces spanned by all subsets of $k$ columns of $\bm G$. A measurement matrix $\bm \Psi$ obeys the restricted isometry property adapted to $\bm G$ with constant $\delta$, if
\[
    (1 - \delta)\Norm{\bm y}_2^2
    \leqslant \Norm{\bm \Psi \bm y}_2^2
    \leqslant (1 + \delta)\Norm{\bm y}_2^2
    \Text{for all} \bm y \in U_k.
\]
We call the smallest $\delta$ for which above chain of inequalities holds the restricted $\ma{G}$-isometry constant ($\ma{G}$-RIC) $\delta_k^{\bm G}$. \qedsymbol
\end{definition}

The following result calculates the $\ma{H}$-RIC of the matrix $\bm A \otimes \bm I_{N_x N_y}$ for an arbitrary $\bm A$ and as such it delivers necessary conditions for efficient, stable and robust recovery to happen when compressing signals with $\bm A \otimes \bm I_{N_x N_y}$ when the signals are sparse with respect to the dictionary $\bm H$ from \eqref{eq:dictionary}.

\begin{theorem}\label{H_G_RIC}
For each $k \in \N$ it holds that the $\ma G$-RIC $\delta_k^{\bm G}$ of the matrix $\bm A$ and the $\bm H$-RIC of $\bm \Psi = \bm A \otimes \bm I_{N_x N_y}$ are equal.
\end{theorem}
\begin{proof}
From the definition of $\delta_{k}^{\bm G}$ and the vectors $\bm \alpha_{n_x,n_y}$ we have for every $(n_x,n_y)$ that
\[
    (1 - \delta_{s_{\rm max}}^{\bm G})\Norm{\bm \alpha_{n_x,n_y}}_2^2
    \leqslant \Norm{\ma{A} \bm \alpha_{n_x,n_y}}_2^2
    \leqslant (1 + \delta_{s_{\rm max}}^{\bm G})\Norm{\bm \alpha_{n_x,n_y}}_2^2.
\]
Together with the definition of $\bm \Phi$ and $\bm y_{n_x,n_y}$ and the properties of the Kronecker product the statement directly follows.
\end{proof}

The theorem above illuminates how, given the proposed sensing scenario, only the properties of the A-scans and their sparse representation influence the recovery performance. Additionally, only the worst A-scan in the sense that it is the least sparse one determines the worst case performance. Now, since randomly sub-selected Fourier matrices are known to have a low $\bm G$-RIC (as outlined below), Theorem~\ref{H_G_RIC} shows how to asses the recovery performance and infer the number of necessary measurements. Moreover, one is able to determine the number of Fourier coefficients $n_f$ such that the $\bm G$-RIP of the appropriate order holds with a high probability, which delivers stable, robust and efficient recovery. This is also formalized in~\cite{candes_DRIP_2010}, where it is shown that
 $\delta_{s_{\rm max}}^{\bm G} < 0.08$
is a sufficient condition for stable recovery to happen in every A-scan. Now assume that we generate $\bm S$ such that it selects a set of rows from $\ten F$ which is drawn from the uniform distribution of all sets of size $n_f$, where
\[
n_f \geqslant
    C s_{\rm max} \delta^{-2} \log^2(1 / \delta) \log N_t \log(s_{\rm max}) \log^2(s_{\rm max} / \delta).
\]
Then, the matrix $\bm S \ten F$ satisfies the standard (not $\ma{G}$) RIP with constant $\delta$ with high probability~\cite{haviv2015fourierrip} for sparsity order $s_{\rm max}$. Additionally,~\cite{krahmer2011ward} shows that if the matrix $\bm S \ten F$ satisfies the RIP of order $k$ with constant $\delta$, the matrix $\bm S \ten F \bm \Sigma$ is a $4\delta$-almost isometry on any set $\mathcal{S}$ of fixed points with $\Abs{\mathcal{S}} = C \exp (k)$, meaning
\[
(1 - 4\delta)(\Norm{\bm a}_2^2)
\leqslant \Norm{\bm S \ten{F} \bm \Sigma \bm x} \leqslant
(1 + 4\delta)(\Norm{\bm a}_2^2)
\]
for all $\bm a \in \mathcal{S}$. This directly implies together with  $\delta_{s_{\rm max}}^{\bm H} < 0.08$ and Theorem~\ref{H_G_RIC} applied for $\bm A = \bm S \ten F \bm \Sigma$ the minimal number of measurements to satisfy
\begin{equation}\label{eq:performance_guarantee}
  n_f \geqslant
  \hat{C} \log N_t s_{\rm max} \log(s_{\rm max}) \log^2(50s_{\rm max}).
\end{equation}
We would like to stress the fact that in the above analysis we derived a performance bound for the 3D reconstruction process, while only dealing with the restricted isometry constants associated to the dictionary of the single dimensional A-scans. In other words, since we are only compressing / subsampling in one dimension, this dimension alone determines the reconstruction performance. Further, the choice of $n_f$ in a practical setup can be based on a worst-case number of defect echoes that are expected to appear in a single A-scan.

\begin{discussion}
\begin{itemize}[wide=0.5em, leftmargin =*, nosep, before = \leavevmode\vspace{-\baselineskip}]
\item On average $\bm \Phi_{\rm rnd}$ selects around $n_f / N_t$ parts of the energy that is contained in the reflected waves. As such of the strategies considered in this work, it has the worst SNR. However due to the random mixing done by the $\bm \Sigma_{n_x, n_y}$ it maximizes the bandwidth of the measurement, since it is approximately the same as the inserted pulse.
\item The strategy using $\bm \Phi_{\rm max}$ maximizes the SNR, since it collects the most energy from the pulse by sampling at the around the peaks in the spectrum. However, due to the typical shape of an ultrasound pulse in time- and frequency domain, the resulting samples are closely spaced, which directly results in a low bandwidth of the acquired signal.
\item As we see in Section~\ref{sec:num_sims}, the two strategies discussed above perform as expected and a trade-off between the two is represented by $\bm \Phi_{\rm nrg}$. So depending on the goal during reconstruction in terms of depth-resolution, the different strategies cover the whole range from high SNR and poor bandwidth to poor SNR and high bandwidth.
\item The problem of estimating the model order $s_{\rm max}$ to set $n_f$ before carrying out the reconstruction is hard to overcome in a \gls{cs} setting. There is a large literature on sparsity order estimation with various advantages and drawbacks, see~\cite{semper2018soe},~\cite{lavrenko2015subnyq},~\cite{lavrenko2015timevarsupp},~\cite{ravazzi2016gammasoe}. Often a satisfactory method for model order selection depends very much on the specific applications' side constraints. In our case, for instance, on the size of a typical defect, the number of defects and their shape.
\item Note for the recovery guarantees that they only represent scaling laws and not explicit bounds on the number of measurements due to the fact that there is a (here) unspecified constant factor involved. However, even if one would compute it explicitly using the results in~\cite{haviv2015fourierrip} and~\cite{krahmer2011ward} these estimates would be too conservative for practical considerations. In case when one has empirical evidence that for a certain sparsity order $s_{\rm max}$, problem size $N$ and number of measurements $n_f$ the recovery is satisfactory, one can infer the necessary number of measurements if the problem size or sparsity change.
\end{itemize}
\end{discussion}


\section{Reconstruction\label{sec:algs}}
\subsection{Algorithms}
Now that we have established the data model and studied the compression schemes, it first is necessary to formulate viable algorithms in order to reconstruct $\bm{a}$ in \eqref{compressed_model} for all discussed compression strategies. Ideally one would solve
\[
  \Min \Norm{\bm a}_0 \Text{s.t.} \bm \Phi \bm H \bm a = \bm b,
\]
which unfortunately is NP-hard~\cite{foucart13mathintcs} in general. So, usually one resorts to solving the convex relaxation of the above problem, which reads as
\[
  \Min \Norm{\bm a}_1 \Text{s.t.} \bm \Phi \bm H \bm a = \bm b.
\]
In this case we end up with a problem that can be solved in polynomial time. Interestingly, there are a number of efficient approximate algorithms that converge rapidly (\gls{ista}, \gls{fista}, \gls{twista}, \gls{stela}) and only rely on computing the matrix vector products $\bm A \cdot \bm x$ and $\bm A^\herm \cdot \bm y$ as their main computational effort. This in turn implies that we end up with well performing algorithms, so called matrix-free algorithms, if we are able to implement the two mappings
\[
    \bm{\phi}_{\ma{A}}: \C^N \rightarrow \C^M, \
    \ma{x} \mapsto \forward{\ma{A}}{\ma{x}} = \ma{A} \cdot \ma{x},
\]
the so-called forward projection, and
\[
    \bm{\beta}_{\ma{A}}: \C^M \rightarrow \C^N, \
    \ma{y} \mapsto\backward{\ma{A}}{\ma{y}} = \ma{A}^\herm \cdot \ma{y},
\]
the so-called backward projection, efficiently for our proposed data model and compression scheme. Specifically, we would have to implement $\bm \phi_{\bm \Phi \bm H}$ and $\bm \beta_{\bm \Phi \bm H}$.

\begin{remark}[\gls{saft}]\label{rem:saft}
It is worth noting that the backward projection applied to the observation data yields the estimate $\ma{a}_{\rm SAFT} = \backward{\ma{\Phi H}}{\ma{y}}$ which is traditionally called \gls{saft} when $\bm \Phi$ is the identity matrix and $h(t) = \delta(t)$ (cf.~for example~\cite{lingvall_jasa_2003}). However, our approach in formulating the data model and compression scheme directly in terms of matrix vector products yields a more general implementation that can also cope with compressed data and more involved physically motivated forward models. As another special case, our formulation also yields a (compressed) synthetic aperture formulation of the \emph{excitelet} reconstruction~\cite{quaegebeur_ultrasonics_2012}.
\end{remark}

In the following we will focus on \gls{fista} as a representative example for many \gls{ssr} algorithms used in compressed sensing, since it provides a good trade-off between simplicity and performance. However, we would like to stress that in general the following considerations provide a blueprint on how to apply matrix-free reconstruction algorithms that use forward and backward projections to the problem at hand.

\begin{algorithm}
	\KwData{Observed measurement vector $\ma{y}$, the initial guess $\bm a_0$\;
	$\bm \phi_{\ma{\Phi H}}(\cdot)$, $\bm \beta_{\ma{\Phi H}}(\cdot)$\ and the largest singular value of $\bm \Phi \bm H$ denoted as $\sigma_{\rm max}$;
}
$\lambda$, $s = 1/\sigma_{\rm max}^2$, $K$\;
$t_1 = 1$, $\ma{\gamma}_1 = \ma{a}_0$, $k=0$\;
\While{$k < K_{\rm max}$}{
$\ma{a}_{k+1}
  = \tau_{\lambda s}(\ma{\gamma}_k - 2 s \bm \beta_{\ma{\Phi H}}(
    \bm \phi_{\ma{\Phi H}}(\ma{\gamma}_k)-\ma{y})
  )$\;
where $\tau_\alpha(\ma{a})_i = {\rm max}\{|x_i|-\alpha, 0\} \cdot {\rm sgn}(x)_i $\;
$t_{k+1} = \frac{1+\sqrt{1 + 4t_k^2}}{2} $\;
$\ma{\gamma}_{k+1} = \ma{a}_k + \left(\frac{t_k - 1}{t_{k+1}}\right)\left(\ma{a}_k - \ma{a}_{k-1}\right)$\;
$k = k+1$\;
}
\caption{Fast Iterative Shrinkage-Thresholding Algorithm (FISTA) \label{alg:FISTA}}
\end{algorithm}

\subsection{Fast Transforms}

To devise efficient implementations for $\bm \phi$ and $\bm \beta$ it is necessary to consider the structure of the matrices $\bm \Phi$ and $\bm H$. In order to start with $\bm H$ we need the following definition.

\begin{definition}[Multilevel Toeplitz Matrices]
Given a $d$-dimensional complex sequence $\ma{t} = [t_{\ma{k}}]$ for the multi index $\ma{k} \in \N^d$, a $d$-level Toeplitz matrix $\bm{T}_{\bm{n},d}$ is recursively defined as
\[\bm{T}_{\bm{n},d}(\bm t) =
\begingroup
\setlength\arraycolsep{1pt}
\begin{bmatrix}
    \bm{T}_{(1,\bm{m}),\ell}        & \bm{T}_{(2 n_1 - 1,\bm{m}),\ell}
    & \dots     & \bm{T}_{(n_1 + 1,\bm{m}),\ell}    \\
    \bm{T}_{(2,\bm{m}),\ell}        & \bm{T}_{(1,\bm{m}),\ell}
    & \dots     & \bm{T}_{(n_1 + 2,\bm{m}),\ell}    \\
    \vdots                          & \vdots
    & \ddots    & \vdots                            \\
    \bm{T}_{(n_1,\bm{m}),\ell}      & \bm{T}_{(n_1 - 1,\bm{m}),\ell}
    & \dots     & \bm{T}_{(1,\bm{m}),\ell}          \\
\end{bmatrix},
\endgroup
\]
where $\bm m = \{n_2, \dots, n_d\}$ and $\ell = d-1$. In this case we say $\bm{T}_{\bm{n},d}(\bm t)$ is a $d$-level Toeplitz matrix generated by $\bm t$.
\end{definition}

For example, consider $d = 2$, $\bm n = (2,2)$, $\bm k = (3,3)$ and $\bm t \in \C^{3 \times 3}$, which yields
\[
\begingroup
\setlength\arraycolsep{1pt}
\bm T_{(2,2),2} =
\begin{bmatrix}
\bm T_{(1,2),1} & \bm T_{(3,2),1} \\
\bm T_{(2,2),1} & \bm T_{(1,2),1}
\end{bmatrix}
=
\begin{bmatrix}
t_{1,1} & t_{1,3} & t_{3,1} & t_{3,3} \\
t_{1,2} & t_{1,1} & t_{3,2} & t_{3,1} \\
t_{2,1} & t_{2,3} & t_{1,1} & t_{1,3} \\
t_{2,2} & t_{2,1} & t_{1,2} & t_{1,1}
\end{bmatrix}.
\endgroup
\]
The essential observation now is that the multiplication of a $d$-level Toeplitz matrix to a vector can be implemented efficiently by means of the fast convolution algorithm by exploiting the $d$-dimensional FFT~\cite{cooley1965FFT,semper_eusipco_2018,kirchhof_ieeeius_2018}. This yields efficient algorithms for $\bm \phi_{\bm{T}_{\bm{n},d}(\bm t)}$ and $\bm \beta_{\bm{T}_{\bm{n},d}(\bm t)}$. Additionally we need to align these generalized Toeplitz matrices in a block matrix structure.

\begin{definition}[Block Multilevel Toeplitz Matrices]
A matrix
\[\bm{B} =
\begingroup
\setlength\arraycolsep{1pt}
\begin{bmatrix}
    \bm{B}_{1,1} & \dots & \bm{B}_{1,n} \\
    \vdots   & \ddots    & \vdots \\
    \bm{B}_{m,1} & \dots & \bm{B}_{m,n} \\
\end{bmatrix},
\endgroup
\]
is called block $d$-level Toeplitz if each $\bm B_{i,j}$ is a $d$-level Toeplitz matrix. We collect the generating elements of $\bm B$ in the $2+d$-dimensional array $\bm b \in \C^{n \times m \times n_1 \times \dots \times n_d}$ such that $\bm b_{i,j} \in \C^{n_1 \times \dots \times n_d}$, $\bm B_{i,j} = \bm{T}_{\bm{n},d}(\bm b_{i,j})$.
\end{definition}

As it turns out, we can show that $\bm H$ in \eqref{compressed_model} is a block Toeplitz matrix. The key observation is that moving any scatterer in the $x$-$y$-plane while also displacing the measurement position by the same amount, does not change the data captured by the transducer up to some boundary effects. This is formalized in the next result.

\begin{theorem}\label{strctH}
The matrix $\bm H \in \R^{N_z N_x N_y \times N_t N_x N_y }$ from \eqref{compressed_model} is block $2$-level Toeplitz, where for the generating elements $\bm h$ it holds that $\bm h \in \R^{N_z \times N_t \times 2 N_x -1 \times 2 N_y -1}$.
\end{theorem}
\begin{proof}
We consider a column $\bm H_{z, x_1, y_1}$ of $\bm H$ with $(x_1,y_1, z) \in
G_{3D}$. If we now pick an arbitrary $(x_2, y_2, ct) \in G_{3D}$, we see that
\[
    [\bm H_{z, x_1, y_1}]_{t, x_2, y_2} = \gamma(z, t, x_1 - x_2, y_1 - y_2),
\]
for some function $\gamma$ depending on the transducer characteristic $g$ and the time of flight $\tau$. And due to the specific structure of $\tau$ we have that $\tau_{x,y}(x_d,y_d,z) = \tau_{0,0}(x_d - x, y_d - y, z)$, so $\bm H$ is $2$-level Toeplitz because of the translational invariance with respect to $x, x_d$ and $y, y_d$. Finally, the asserted structures of $\bm H$ and $\bm h$ follow easily.
\end{proof}

Note that above result implicitly also states how the defining array $\bm h \in \C^{N_t \times N_z \times 2 N_x -1 \times 2 N_y -1}$ to generate the block 2-level Toeplitz Matrix $\bm H$ should be computed.

Now, by accounting for the block Toeplitz structure, we get $\bm \phi_{\bm H}$ and $\bm \beta_{\bm H}$ by noticing that
\[
  \bm H \cdot \bm a = \left[
    \sum_{j = 1}^{N_t} \bm H_{i,j} \bm a_j
  \right]_{i = 1}^{N_z}
  = \left[
    \sum_{j = 1}^{N_t} \bm \phi_{\bm H_{i,j}}(\bm a_j)
  \right]_{i = 1}^{N_z},
\]
where the $\bm H_{i,j}$ are $2$-level Toeplitz and each $\bm a_j$ is a subvector of $\bm a$ of size $N_x N_y$. However, consider the specific example from Section~\ref{sec:motivation} where we take measurements on a scanning grid of size $100 \times 100$ and measurement has $1000$ time samples. Then, assuming that the reconstruction grid is identical to the measurement grid, the $1000^2$ generating elements each have size $199 \times 199$, which would require $\approx \SI{158}{GB}$. So, even for moderately sized problems and while exploiting the inherent Toeplitz structure, we cannot fit the generating sequences of each $\bm H_{i,j}$ into memory at once. In these cases the generating elements have to be recomputed during each transformation step. To this end, we propose to calculate each $\bm h_{i,j}$ on the fly during the multiplication of $\bm H$, see Algorithm~\ref{alg:transform} for the details. This is especially beneficial when it is carried out on the GPU.

Next, we proceed with the analysis of the structure of the compression matrix $\bm \Phi$ in order to devise $\bm \phi_{\bm \Phi}$ and $\bm \beta_{\bm \Phi}$. In its most general form we have by \eqref{general_compression} that
\[
  \bm \Phi = \bm\BlkDiag\{\bm \Phi_{1,1}, \dots, \bm \Phi_{N_x,N_y}\},
\]
i.e., the matrix $\bm \Phi$ is a block-diagonal matrix, where each block consists of a product
\[
  \ma{S}_{n_x, n_y} \ten{F} \bm \Diag(\bm \xi_{n_x, n_y}).
\]
In terms of matrix vector products, this means that multiplication with $\bm \Phi$ is a blockwise procedure, where each block is processed first by a pointwise multiplication ($\odot$) with $\bm \xi_i$, an FFT and lastly a subselection of the vector in the frequency domain. Thus, it is trivial to implement these matrix vector products efficiently. Finally, we note that $\bm \phi_{\bm \Phi \bm H} = \bm \phi_{\bm \Phi} \circ \bm \phi_{\bm H}$ and $\bm \beta_{\bm \Phi \bm H} = \bm \beta_{\bm H} \circ \bm \beta_{\bm \Phi}$, where $\circ$ denotes function concatenation. So given two distinct implementations for the matrix vector products for $\bm \Phi$ and $\bm H$ we only have to concatenate them.

\begin{algorithm}
\SetAlgoLined
\KwData{Input data $\bm x \in \C^{N_z \times N_x \times N_y}$}
\KwResult{Transformed data $\bm{\phi}_{\bm \Phi \bm H}(\bm x) = \bm y \in \C^{n_f \times N_x \times N_y}$}
Zeropad $\bm x$ to $\bm x_0 \in \C^{N_z \times 2 N_x - 1 \times 2 N_y - 1}$\;
Apply a $2$D-FFT along dimensions $2$ and $3$ to $\bm x_0$ to get $\hat{\bm x}_0$\;
Set $\hat{\bm y}_0 = \bm 0 \in \C^{N_z \times 2 N_x - 1 \times 2 N_y - 1}$\;
\For{$i_t = 1, \dots, N_t$}{
  \For{$i_z = 1, \dots, N_z$}{
    Calculate $\bm h_{i_t, i_z} \in \C^{2 N_x - 1 \times 2 N_y - 1}$\;
    Apply a $2$D-FFT to $\bm h_{i_t, i_z}$ to get $\hat{\bm h}_{i_t, i_z}$\;
    $\hat{\bm y}_{0,i_t} \mathrel{{+}{=}} \hat{\bm h}_{i_t, i_z} \odot \hat{\bm x}_{0,i_t}$\;
  }
}
Apply a $2$D-iFFT to $\hat{\bm y}_{0}$ along dimensions $2$ and $3$ to get $\bm y_{0}$\;
Revert the zero-padding on $\bm y_0$ to get $\bm y \in \C^{N_z \times N_x \times N_y}$\;
\For{$i_x = 1, \dots, N_x$}{
  \For{$i_y = 1, \dots, N_y$}{
    $\bm y_{\cdot, i_x, i_y} \leftarrow \bm \xi_{i_x, i_y} \odot \bm y_{\cdot, i_x, i_y}$\;
  }
}
Apply a $1$D-FFT along the first dimension to get $\hat{\bm y} \in \C^{N_z \times N_x \times N_y}$\;
\For{$i_x = 1, \dots, N_x$}{
  \For{$i_y = 1, \dots, N_y$}{
    $\bm y_{\cdot, i_x, i_y} \leftarrow \bm S_{i_x, i_y} \hat{\bm y}_{\cdot, i_x, i_y} $\;
  }
}
Return $\bm y$.
\caption{Computation of $\bm \phi_{\bm \Phi \bm H}$}
\label{alg:transform}
\end{algorithm}

Algorithm~\ref{alg:transform} schematically displays how to carry out the multiplication with $\bm \Phi \bm H$ efficiently. The algorithm for $(\bm \Phi \bm H)^\herm$ can be derived in a similar manner.

\begin{discussion}
\begin{itemize}[wide=0.5em, leftmargin =*, nosep, before = \leavevmode\vspace{-\baselineskip}]
\item It is worth noting that Algorithm~\ref{alg:transform} should be implemented in a blocked manner. This means that the loops for $i_x$ and $i_y$ should be replaced with loops over blocks containing several $i_{x_i}, \dots i_{x_i + b}$ and $i_{y_i}, \dots i_{y_i + b}$ at once, since most high-level programming languages allow for faster processing of these blocks, especially when working on a GPU. This also steers the amount of system memory the transformation occupies and can be tuned to the system specifications at hand.
\item In the extreme case when one has enough system memory available to store $\bm h$ entirely, one should do so in order to maximize performance.
\item Additionally, the problem dimensions $N_x$ and $N_y$ do not influence the \gls{fft} performance that much, since for badly conditioned \gls{fft} sizes in terms of prime factors, once can exploit Bluestein's algorithm~\cite{bluestein1970bluestein}, which expresses the Fourier transform as a cyclic convolution where one can use zero-padding. Most modern \gls{fft} implementations have heuristics in place to decide whether to use this alternative approach or not.
\end{itemize}
\end{discussion}

\subsection{The Largest Singular Value}

As we see in Algorithm~\ref{alg:FISTA} it is necessary to compute or at least estimate $\sigma_{\rm max}$ for the matrix $\bm \Phi \bm H$. When estimating the singular value, one should take care that it is not underestimated, because in this case one expects the objective function minimized by \gls{fista} to be smoother than it actually is. This in turn leads to more aggressive (and in this case unjustified so) iteration steps, which ultimately results in divergence of the algorithm. Consequently it is better to overestimate $\sigma_{\rm max}$ to avoid divergence at the cost of slower convergence.

However, in a matrix-free setting this is no trivial endeavor, since algorithms that rely on having the whole matrix in system memory cannot be used.

\subsubsection{Backtracking}

A first approach would be to use a backtracking version of \gls{fista}~\cite{beck_siam_2009}, which does not need the singular value as an input, but rather a crude lower bound $\sigma_{\rm low}$ together with a scaling parameter $\eta > 1$. Then, in each step one determines an appropriate stepsize based on a local estimate $\sigma_{\rm loc}$ within
\[
  \sigma_{\rm low}
  \leqslant \sigma_{\rm loc}
  \leqslant \eta \cdot \sigma_{\rm max}.
\]
Ultimately, $\eta$ influences the speed of convergence such that higher values of $\eta$ result in slower convergence due to the overestimation of the singular value. Additionally, this comes at the cost of evaluating the objective function and a suitable quadratic approximation numerous times during the iteration, leading to an impractical amount of computations that are necessary for convergence, since we have to call $\bm \phi_{\bm \Phi \bm H}$ twice per backtracking step. Hence, in large scale scenarios it might be advantageous to have means for acquiring the largest singular value. In the following we present two alternatives.

\subsubsection{Estimation}

Another approach would be to find a suitable estimate $\hat{\sigma}_{\rm max}$ that is cheap to compute, which still allows proper convergence~\cite{kirchhof_ieeeius_2018}. In this case one has to ensure that the estimated singular value is bounded from below by the true $\sigma_{\rm max}$. For the product $\bm \Phi \bm H$ at hand we proceed as follows.

A simple bound for the largest singular value is given by
\[
  \frac{1}{\sqrt{N}}\Norm{\bm \Phi \bm H}_F
  \leqslant \sigma_{\rm max}
  \leqslant \Norm{\bm \Phi \bm H}_F,
\]
which would imply to estimate $\hat{\sigma}_{\rm max} = \Norm{\bm \Phi \bm H}_F$, which in terms of convergence provides poor results, since it forces \gls{fista} to take too conservative steps. As such, we propose to use the geometric mean of the upper and lower bound
\begin{equation}\label{sigma_hat}
  \hat{\sigma}_{\rm max} = \frac{\Norm{\bm \Phi \bm H}_F}{N^{(1/4)}},
\end{equation}
for which it is necessary to compute $\Norm{\bm \Phi \bm H}_F$. However, since we have no direct access to the entries of $\bm \Phi \bm H$ we need to estimate this as well. First we notice that
\[
  \Norm{\bm \Phi \bm H}_F^2 = \sum_{z_d} \sum_{x_d} \sum_{y_d} \Norm{\bm \Phi \bm H_{z_d, x_d, y_d}}_2^2.
\]
this means we are summing over the squared norms of all atoms, where each atom belongs to a single defect position $(x_d, y_d, z_d)$. Now, neglecting boundary effects within the individual atoms we can simplify to
\[
  \Norm{\bm \Phi \bm H}_F^2 \approx
    N_x N_y \sum_{z_d} \Norm{\bm \Phi \bm H_{x_d^0, y_d^0,z_d}}_2^2
\]
for some representative target position $(x_d^0, y_d^0, \cdot)$. This leaves us with calculating the innermost summand for which we have
\begin{align*}
  \Norm{\bm \Phi \bm h(x_d^0, y_d^0,z_d)}_2^2
  \approx \sum_{n_x = 1}^{N_x} \sum_{n_x = 1}^{N_x}
  g_{n_x \Delta x,n_y \Delta y}(x_d^0,y_d^0, z_d)^2 \\
  \Norm{
  \bm \Phi_{n_x,n_y}
  \bm h_{n_x,n_y} (x_d^0, y_d^0,z_d)
  }^2.
\end{align*}
Finally, depending on the actual subselection strategy employed by $\bm \Phi$ we can approximate the largest singular value by the above estimate of the Frobenius norm.

\subsubsection{Iterative Approximation}

To circumvent possible problems from an inaccurate estimate of the singular value, iterative algorithms~\cite{Lehoucq1996arnoldi} to approximate it have been developed and finally implemented in \gls{arpack}~\cite{lehoucq1998arpack}. These again only rely on $\bm \phi_{\bm \Phi \bm H}$ and $\bm \beta_{\bm \Phi \bm H}$ and for a specific scenario $\bm \Phi \bm H$ one can cache this approximate result. However one cannot guarantee that the singular value is not under-estimated in magnitude.

\begin{discussion}
\begin{itemize}[wide=0.5em, leftmargin =*, nosep, before = \leavevmode\vspace{-\baselineskip}]
  \item In cases where the evaluation of $\bm \phi_{\bm \Phi \bm H}$ takes up to several minutes it is not advisable to use a backtracking scheme in \gls{fista}. Instead one should use these additional projections during an Arnoldi iteration and approximate and store the largest singular value directly.
  \item As already noted, one can store the results from \gls{arpack} for later reuse and avoid the repeated calculation of the largest singular value.
  \item Additionally, the approach we took for the estimation of $\sigma_{\rm max}$ using \eqref{sigma_hat} can easily be generalized and altered to account for different data and compression models.
  \item Although one cannot guarantee that \gls{arpack} estimates $\sigma_{\rm est} > \sigma_{\rm max}$ it provides a tolerance $\delta_{\sigma} \geqslant \vert \sigma_{\rm est} - \sigma_{\rm max} \vert$. So ultimately one can ensure that $\sigma_{\rm max} \leqslant \sigma_{\rm est} + \delta_{\sigma}$, which can in turn be used in \gls{fista} safely.
\end{itemize}
\end{discussion}


\section{Implementation considerations\label{sec:implementation}}
In the following section, we analyze the proposed sampling and modeling strategies from Secs.~\ref{sec:data_model}~and~\ref{sec:fourier_subsampling} in terms of their implementation effort in hardware as well as computation complexity.
Aside from the final measurement quality, implementation effort is a crucial aspect when choosing an architecture for a particular measurement application.
The three-stage model introduced in Section~\ref{sec:intro} offers an intuitive abstraction, well suited for the following discussion:

To recap, the first \emph{Data management} stage comprises of first, the \emph{acquisition} frontend, generating a \emph{raw data} stream of digital representations for the analogue pulse-echo signals, and second, the interfaces for \emph{Data handling and storing} into non-volatile memory banks.
Representing the \emph{sense making} stage, an off-site \emph{computation unit} analyzes the stored \emph{raw data} and distills it into interpretable information.
The final \emph{Decision making} stage will not be discussed, since it bears no relevance to implementation aspects.

Based on a common example scenario, Table~\ref{tab:implementation_parameters} summarizes key design parameters and performance indications when comparing the following three measurement architectures:
\begin{enumerate}[label=(\Roman*)]
  \item\label{sys:SAFT} A state-of-the-art system based on oversampling and \gls{saft}, cf. Remark~\ref{rem:saft}.
  \item\label{sys:Model} A critically sampling system using the physically motivated forward model $\ma{H}$ defined in this work (cf.~Sec.~\ref{sec:data_model}).
  \item\label{sys:CS_model} A sub-Nyquist sampling system extending~\ref{sys:Model} by the sub-Nyquist sampling strategies $\ma{\Phi}$ from Sec.~\ref{sec:fourier_subsampling}.
\end{enumerate}

A single architecture, satisfying the widely diverse and often conflicting requirements for the set of all measurement applications, cannot be found optimally.
Some handheld measurement units require battery operation on mobile network connections, while other units, irreplaceably built into long-lasting structures, deliver their raw data over wired networks.
These conflicting demands require making compromises between cost, size, power- and energy usage, ruggedness, data path or accuracy.

Notably, the major benefit of our proposed compressive architecture over common~\gls{saft} implementations (aside from image quality) is that it allows trading the amount of collected raw data against computation complexity later in the process.
In choosing the number of obtained Fourier coefficients per A-scan (in Sec.~\ref{sec:num_sims} we show that already one is sufficient), our proposed compressed architecture relates the raw data rate to the actual amount of relevant signal information, rather than some artificial grid constraint (as is the case for oversampling \gls{saft}).
This makes it especially useful for applications suffering from \emph{data handling} bottlenecks~\cite{cawley_structural_health_monitoring}.

\begin{table}[tb!]
  \makebox[\textwidth][c]{
    \begin{minipage}{0.99\textwidth}
      \begin{tabular}{@{\rule[-1ex]{0pt}{4ex}} m{-3pt} r|c@{\hspace{1em}}c@{\hspace{1em}}c}
      \parbox[t]{2mm}{\multirow{3}{*}{\rotatebox[origin=c]{90}{\footnotesize \textit{Scenario}}}}%
        & Volume Surface%
            & \multicolumn{3}{c}{all methods $100\!\times\!100$ Points}%
            \\[-1ex]
        & Max. round-trip-time%
            & \multicolumn{3}{c}{all methods~\SI{25}{\micro\second}}%
            \\[-1ex]
        & Pulse bandwidth%
            & \multicolumn{3}{c}{all methods~\SI{10}{\mega\hertz}}%
            \\
      \hline
      \hline
      &Implemented system & \ref{sys:SAFT} & \ref{sys:Model} & \ref{sys:CS_model} \\[-1ex]
      &  & \gls{saft} & $\bm{H}$ & $\bm{\Phi H}$\\[0.5ex]
      \hline
      \parbox[t]{2mm}{\multirow{6}{*}{\rotatebox[origin=c]{90}{\footnotesize \textit{Data Management}}}}%
        & ADC (\SI{16}{\bit}) rate [\si{\sample\per\second}]%
          & $k_O\cdot\SI{20}{M}$%
          & \SI{20}{M}%
          & {\boldmath $\leqslant\!\SI{20}{M}$}$^{\ast}$%
          \\
        & ADC data stream [\si{\bit\per\second}]%
          & $k_O\cdot\SI{320}{M}$%
          & \SI{320}{M}%
          & {\boldmath $\leqslant\!\SI{320}{M}$}$^{\ast}$%
          \\
        & Samples per A-scan $N_t$%
          & $k_O\cdot\num{500}$%
          & \num{500}%
          & {\boldmath $\geqslant\!\num{1}$}%
          \\
        & A-scan data size$^{\ast\ast}$%
          & $k_O\cdot\SI{1}{\kilo\byte}$%
          & \SI{1}{\kilo\byte}%
          & \textbf{\SI{8}{\byte}}%
          \\
        & Volume data size$^{\ast\ast}$%
          & $k_O\cdot\SI{10}{\mega\byte}$%
          & \SI{10}{\mega\byte}%
          & \textbf{\SI{0.08}{\mega\byte}}%
          \\[-1ex]
        & \dots as normalized ratio$^{\ast\ast\ast}$%
          & $\approx10^{3.3}$%
          & $\approx10^{2.1}$%
          & {\boldmath $1$}%
          \\[1ex]
      \hline
      \hline
      \parbox[t]{2mm}{\multirow{4}{*}{\rotatebox[origin=c]{90}{\footnotesize \textit{Sense making}}}}%
        & Computation Complexity%
          & {\boldmath $+$}%
          & $\circ$%
          & $-$%
          \\[-1ex]
        & \dots as normalized ratio$^{\ast\ast\ast}$%
          & {\boldmath $1$}%
          & $\approx10^{2.8}$%
          & $\approx10^{4.8}$%
          \\
        & Depth-axis resolution%
          & fixed%
          & \textbf{variable}%
          & \textbf{variable}%
          \\[-1ex]
        & Considers physical model%
          & no%
          & \textbf{yes}%
          & \textbf{yes}%
          \\[-1ex]
        \multicolumn{2}{@{\rule[-1ex]{0pt}{4ex}} r|}{\quad Approximates inverse problem}%
          & no%
          & no%
          & \textbf{yes}%
          \\[-1ex]
        & Point focus quality%
          & $-$%
          & {\boldmath $+$}%
          & $(\circ \ldots +)$%
          \\[1ex]
      \end{tabular}
      \vspace{4pt}
    \end{minipage}\\%
  }
  \vspace{4pt}
  {%
    \footnotesize
    \textit{$k_O =$ Oversampling factor ; $k_F =$ \gls{fista} iterations}\\
    $^{\ast}$ \textit{equal to when using \gls{fft} on raw samples, $<$ when employing~\cite{mulleti_ieeetsp_2020}}\\
    $^{\ast\ast}$ \textit{Data storage, complex single-precision float (8~byte)}\\
    $^{\ast\ast\ast}$ \textit{Relative, given for $k_O\!=\!16$ and $k_F\!=\!50$ and $\Delta z\!=\!t_s\!\cdot\!c$.}\\%
  }
  \caption{Comparing implementation effort and performance of the system~\ref{sys:CS_model} proposed in this work compared to the two example systems~\ref{sys:SAFT} and~\ref{sys:Model}.
  \label{tab:implementation_parameters}}
  \vspace{-2mm}
\end{table}
\subsection{Effort for data acquisition\label{sec:implementation_data_acquisition}}
For the implementation of the signal acquisition, multiple strategies exist, where each has individual benefits and drawbacks (depending on the application).
Following, a general overview is provided to allow for good architectural decisions.
\subsubsection{Direct time-domain sampling}\label{sampling:time}
A standard \gls{adc} linearly represents the time-domain pulse-echo response signal as a digital vector.
To fully represent all signal information in the raw data stream, the sampling rate $f_s$ must be at least the critical rate $f_{\rm crit}\!=\!2f_{\rm max}$, where $f_{\rm max}$ is the highest signal frequency component.
To keep computation cost low, a simple delay-based \gls{saft} model (disregarding more complex propagation effects) is commonly chosen.
By introducing an oversampling factor $k_O\!=\!\sfrac{f_s}{f_{\rm crit}}$, good depth-resolution with acceptable visual artifacts is achieved \cite{lingvall_jasa_2003}.
For common choices like $k_O\!=\!16$ the amount of added redundant data is immense.

This work's generalized matrix formulation $\bm H$ adds an improved physical propagation model, increasing visual quality considerably at the price of higher computation cost during the \emph{sense making} stage (see Sec.~\ref{sec:data_processing}).
However, data amount is reduced substantially, since $k_O\!=\!1$ can always be chosen.

Albeit very poor information efficiency and high demands on \gls{adc} frontend and \emph{data handling}, oversampling \gls{saft} may still be advantageous when available computation resources are limited, i.e. for battery-powered on-site inspection systems.
\subsubsection{Digital Fourier coefficient sampling}\label{sampling:fourier_digital}
Additionally, we can extend~\ref{sys:Model} by a sub-sampling/compression matrix to form the \gls{cs} system~\ref{sys:CS_model}, allowing for source-compression by keeping only a few Fourier coefficients of each A-scan.
In reconstructing an image from all A-scans jointly, spatial redundancy is exploited such that keeping as low as one coefficient per A-scan is sufficient in many scenarios.
While this greatly improves \emph{Data handling}, the computation-intense algorithms from section~\ref{sec:algs} must now be applied in the \emph{sense making} stage.
The coefficients can be computed directly from the \gls{adc} output using the \gls{fft} or the Goertzel algorithm.
In this case the implementation effort up to the \gls{adc} is identical to the critical sampling case in~\ref{sampling:time}.
This approach works best when collected data is processed off-site, where bulk computation power is readily available, and the measurement device can afford to compute some Fourier coefficients on-the-fly.
\subsubsection{Analogue Fourier coefficient sampling}\label{sampling:fourier domain analogue}
Retrieving the Fourier coefficients for system~\ref{sys:CS_model} can also can be implemented using sub-Nyquist sampling, as proposed in~\cite[Sec.~3]{mulleti_ieeetsp_2020}.
This requires the signal to be filtered by a Sum-of-Sincs filter in analog domain, which is sparsely sampled with an \gls{adc} operating below the critical sampling rate.
From the collected samples, the desired Fourier coefficients (denoted as the set $\mathcal{K}_n$ in~\cite[Sec.~3]{mulleti_ieeetsp_2020}) are retrieved by solving a system of linear equations.
Multi-channel setups, such as~\cite{perez_ieeeius_2019}, may easily be supported by moving from sparse- to interleaved sampling of multiple analogue channels by combining a single \gls{adc} with a multiplexer.
This makes this approach attractive for low-power single- or multichannel applications.
Further combined with integrated \gls{cmos} technology, low-cost and low-footprint multi-channel frontends can be achieved.
This allows for the measurement strategies $\ma{\Phi}_{\rm nrg}$ and $\ma{\Phi}_{\rm max}$ to be implemented. To realize $\ma{\Phi}_{\rm rnd}$ the multiplication of $\ma{\Sigma}_i$ additionally needs to be implemented in the analog domain.

%
%
\subsection{Effort for data processing (computation complexity)\label{sec:data_processing}}
In the \emph{Sense making} stage, the raw data is condensed to interpretable information by a computation unit, where  computation effort is a crucial parameter.
Since the main computing effort for \gls{saft}, \gls{fista} and \gls{arpack} lies in $\bm \phi_{\bm \Phi \bm H}$ and $\bm \beta_{\bm \Phi \bm H}$, we study the relative complexity of Algorithm~\ref{alg:transform} in relation to the spatial grid sizes $N_x$, $N_y$ and $N_t\!=\!N_z$.
The cost of $\bm \Phi \cdot \bm x$ (for some vector $\bm x$) is dominated by the cost of the $N_x\!\cdot\!N_y$ \gls{fft}s, yielding a complexity of $\mathcal{O}(N_t \log(N_t) N_x N_y)$.
For $\bm H$, we see that we have to compute the generating elements\footnote{Note that this can be pre-computed if $\ma{H}$ fits into the memory. If they need to be computed on-the-fly, the computation complexity of the generating elements using the model from Sec.~\ref{sec:us_data_model} is neglibible compared to the computation of the actual matrix-vector product.} in $\mathcal{O}(N_t^2 N_x N_y)$ and then compute all $2$D convolutions in $\mathcal{O}(N_t N_z N_y \log(N_y) N_x \log(N_x))$.
The total computation complexity for the \gls{saft} (cf.~Remark~\ref{rem:saft}) matrix-vector product $\bm H \cdot \bm x$ (which is equal to the complexity of the product $\bm \Phi \bm H \cdot \bm x$) is $\mathcal{O}(S)=\mathcal{O}(N_t N_z N_y \log(N_y) N_x \log(N_x))$.
Calculating $\sigma_{\rm max}$ requires $k_A$ \gls{arpack} iterations with a complexity of $\mathcal{O}(k_A S)$ each.
Since $\sigma_{\rm max}$ may be reused, the complexity for one reconstruction is $\mathcal{O}(k_F S)$, scaling only with the $k_F$ iterations of \gls{fista}.

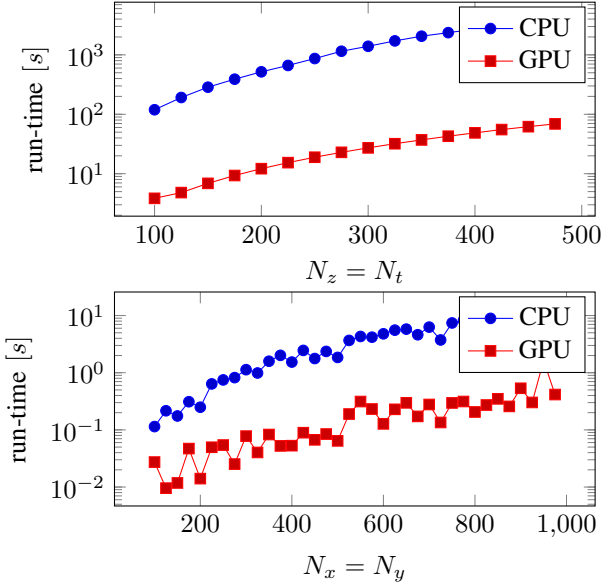
\begin{figure}
\begin{tikzpicture}
\begin{axis}[
  name=z,
  width=0.9\linewidth,
  height=0.5\linewidth,
  xlabel={$N_z = N_t$},
  ylabel={run-time $[s]$},
  ymode=log,
  ]
  \addplot+ table {bench_results/nz_CPU.txt};
  \addplot+ table {bench_results/nz_GPU.txt};
  \legend{CPU, GPU};
\end{axis}
\begin{axis}[
  name=xy,
  width=0.9\linewidth,
  height=0.5\linewidth,
  at=(z.below south),
  anchor=above north,
  xlabel={$N_x = N_y$},
  ylabel={run-time $[s]$},
  ymode=log,
  ]
  \addplot+ table {bench_results/nxy_CPU.txt};
  \addplot+ table {bench_results/nxy_GPU.txt};
  \legend{CPU, GPU};
\end{axis}
\end{tikzpicture}
\caption{Empirical study of the influence of the scene dimensions on the run-time of $\bm \phi_{\bm \Phi \bm H}$. For varying $N_z$ we choose $N_x = N_y = 100$, for varying $N_{x,y}$ we choose $N_z=5$.}
\label{fig:runtime}
\end{figure}

We carried out an empirical study on the influence of the grid sizes in $z$-direction independently from $x$ and $y$ in Figure~\ref{fig:runtime}.
As expected, the transform scales quadratically with $N_z$ and $N_t$.
Varying $N_x$ or $N_y$ influences the \gls{fft} size, exhibiting some ripple in the run-time plots of Fig.~\ref{fig:runtime}, depending on the prime factorization of $2N_x - 1$ and $2N_y - 1$.

In case of the simple delay-based oversampling system~\ref{sys:SAFT} (requiring no convolutions, see section~\ref{sampling:time}), a very low computation complexity of $\mathcal{O}(k_O N_z N_y N_x)$ is attained.
Compared to this model, the complexity of system~\ref{sys:Model} is higher by a factor of $\mathcal{O}(\sfrac{N_t}{k_O}\log(N_y)\log(N_x))$.
In the case of system~\ref{sys:CS_model}, $\bm{\Phi H}$ is applied twice in each \gls{fista} iteration, increasing the computation effort by another factor of $2k_F$.
In exchange, system~\ref{sys:CS_model} actually approximates the inverse problem, as opposed to the other systems, that solely perform projections.

In total, Table~\ref{tab:implementation_parameters} reports the total computation complexity to be $\approx10^3$ times higher for~\ref{sys:Model} and $\approx10^5$ times higher for~\ref{sys:CS_model} compared to simple oversampling \gls{saft} for the example scenario.
However, this comes along with a significant reduction of total collected raw data amount and an improvement of reconstruction quality (see e.g., Fig.~\ref{fig:muse_zoom}).

\begin{discussion}
  \begin{itemize}[wide=0.5em, leftmargin =*, nosep, before = \leavevmode\vspace{-\baselineskip}]
    \item For the example setting, Table~\ref{tab:implementation_parameters} reports a data reduction of~\SI{\approx 99}{\percent} when compared against system~\ref{sys:Model} and~\SI{\approx 99.9}{\percent} when compared against system~\ref{sys:SAFT}, solely from reducing spatial redundancy by implementing~\ref{sys:CS_model}.
    \item When applying the physically motivated forward model $\ma{H}$, it is possible to choose the depth-axis resolution $\Delta z$ independent of the particular sampling rate $f_S$. This is opposed to the state-of-the-art oversampling \gls{saft} system, where a high $f_S$ is required to achieve small $\Delta z$, and as a consequence, measurement data must be reacquired.
    \item The implementation of the strategies $\ma{\Phi}_{\rm nrg}$ and $\ma{\Phi}_{\rm max}$ is slightly easier compared to $\ma{\Phi}_{\rm rnd}$, especially terms of hardware components in the analog domain, since they do not require the additional element-wise multiplication due to $\ma{\Sigma} = \ma{I}$.
    \item The benefits of the proposed \gls{cs}-architecture~\ref{sys:CS_model}, amongst which is that it actually approximates the inverse problem, appear to surpass the large computation complexity, considering the recent advances in \gls{cpu} and \gls{gpu} performance.
  \end{itemize}
\end{discussion}


\section{Asymptotic performance \label{sec:asymptotic_performance}}
\begin{figure}
    \input{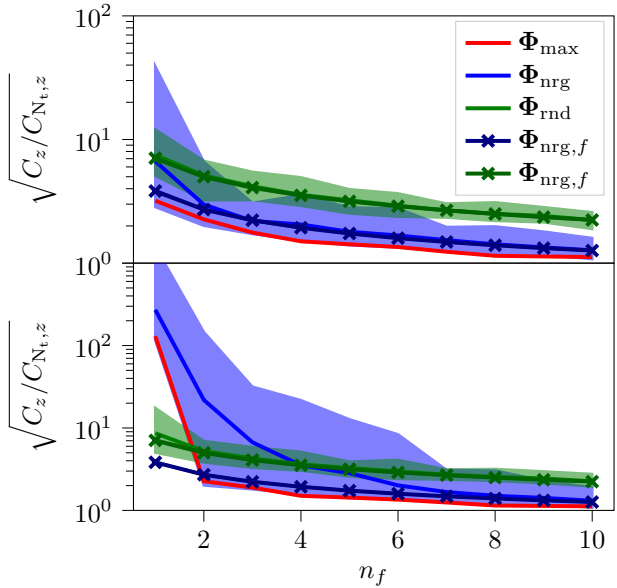}
    \caption{Asymptotic performance of the different subsampling strategies for varying $n_f$.\label{fig:crb_results}}
\end{figure}

As one of the main questions in \gls{ndt} is how accurately we can localize a defect within a specimen, we investigate this accuracy in terms of the \gls{crb} for a single scatterer. In particular, we investigate the loss in localization accuracy as a function of the number of Fourier coefficients per A-scan $n_f$ comparing the proposed sub-sampling strategies. The \gls{crb} is computed by means of the \gls{fim}, which in turn depends on our assumed noise statistics. Assuming $\ma{n}$ in \eqref{compressed_model} to be zero-mean circularly symmetric white complex Gaussian noise with variance $\sigma^2$, it follows that $\ma{y} \sim {\mathcal{CN}}(\ma{\Phi b}, \sigma^2 \ma{I})$. The measurements depend on the set of parameters
\begin{equation*}
    \ma{u} = [\ma{p}^\trans, a, \varphi, \sigma_n^2]^\trans \in \R^{6}
\end{equation*}
where $\ma{p} = [x_1, y_1, z_1]^\trans \in \R^{3}$ comprises the scatterer location, $a$ is the scatterer amplitude and $\varphi$ is the scatterer phase, such that in \eqref{eq:a_scanmodel} we would get $a_1 = a \textrm{e}^{\jmath \varphi}$. We further assume a Gaussian pulse with
\begin{equation*}
    h(t) = {\rm e}^{-\alpha^2 (t-\tau_{x,y})^2 + \jmath \omega_c (t- \tau_{x,y})},
\end{equation*}
and a transducer directivity of the form
\begin{equation*}
    g(x_d, y_d, z_d) = {\rm e}^{\frac{-((x_d - x)^2 + (y_d - y)^2)}{(\tan(\theta) z_d^2)^2}},
\end{equation*}
where $\omega_c$, $\alpha$ and $\theta$ are perfectly known.
Note that the covariance of $\ma{y}$ is independent of $\ma{p}^\trans$, $a$, and $\varphi$ while $\ma{\Phi b}$ is independent of $\sigma_n^2$. Further, $\ma{\Phi}$ is independent of $\ma{u}$. Since $\ma{y}$ follows a Gaussian distribution and together with the previous observations, the Slepian-Bangs formulation~\cite{slepian} of the FIM is given by
\begin{equation}\label{eq:fim}
   \ma{J} = \frac{2}{\sigma_n^2} \text{Re}\left\{\left(\frac{\partial \ma{b}}{\partial \ma{u}^\trans}\right)^\herm \ma{\Phi}^\herm \ma{\Phi}\frac{\partial \ma{b}}{\partial \ma{u}^\trans} \right\} \in \R^{5 \times 5}.
\end{equation}
Let $\ma{\hat H}(\ma{\omega}) =\Diag\{{\rm e}^{-\jmath \ma{\omega}\tau_{x,y}(x_d, y_d, z_d))}\} \odot \ma{H}(\ma{\omega}) $. With this definition at hand, the elements of the \gls{fim} can be computed by the blocks of $\ma{\Phi} \frac{\partial\ma{b}}{\partial\ma{u}^\trans}$ given by
\begin{align}
    \ma{S}_i \ten{F} \ma{\Sigma}_i \ten{F}^\herm \frac{\partial \ma{b}_i}{\partial \ma{p}} &= a_1 \ma{S}_i \ten{F} \ma{\Sigma}_i \ten{F}^\herm \left(\ma{\hat H}(\ma{\omega})  \odot  \right.\\ \nonumber
    &\quad \left. \left( \frac{\partial\ma{g}}{\partial \ma{p}} - \jmath \ma{\omega} \frac{\partial\tau_i(x_1,y_1,z_1)}{\partial \ma{p}} \right)\right) \\
    \ma{S}_i \ten{F} \ma{\Sigma}_i \ten{F}^\herm \frac{\partial \ma{b}_i}{\partial a} &= \ma{S}_i \ten{F} \ma{\Sigma}_i \ten{F}^\herm\frac{\ma{b}_i}{a} \\
    \ma{S}_i \ten{F} \ma{\Sigma}_i \ten{F}^\herm \frac{\partial \ma{b}_i}{\partial \varphi} &= \jmath \ma{S}_i \ten{F} \ma{\Sigma}_i \ten{F}^\herm \ma{b}_i
\end{align}
with
\begin{align*}
    \frac{\partial \tau_i(x_1, y_1, z_1)}{\partial x_1} &= \frac{c_0}{2} \frac{(x_1 - x_i)}{\sqrt{(x_1-x_i)^2 + (y_1 - y_i)^2 + z_1^2}}\\
    \frac{\partial \tau_i(x_1, y_1, z_1)}{\partial z_1} &= \frac{c_0}{2} \frac{z_1}{\sqrt{(x_1-x_i)^2 + (y_1 - y_i)^2 + z_1^2}} \\
    \frac{\partial g_i{x_1, y_1, z_1}}{\partial x_1} &= -2\frac{x_1 - x_i}{(z_1\tan(\alpha))^2} \cdot g_i(x_1, y_1, z_1) \\
    \frac{\partial g_i{x_1, y_1, z_1}}{\partial z_1} &= -\frac{(x_1 - x_i) ^2 - (y_1 - y_i)^2}
    {z_1^3\tan(\alpha)^2}\cdot g_i(x_1, y_1, z_1).
\end{align*}
The derivatives with respect to $y_1$ follow from the derivatives with respect to $x_1$ in a similar fashion as above. The \gls{fim} can be formulated as a block matrix by separating the nuisance parameters, leading to
\begin{equation}\label{eq:block_fim}
    \ma{J}= \frac{2}{\sigma_n^2} \begin{bmatrix}
        \ma{J}_{\ma{p}} & \ma{J}_{\ma{p}, (a,\varphi)} \\
        \ma{J}_{\ma{p}, (a,\varphi)}^\trans & \ma{J}_{a, \varphi}
    \end{bmatrix}.
\end{equation}
The \gls{crb} for the defect location is then given by
\begin{equation}
    \ma{c} = [C_x, C_y, C_z]^\trans = \Diag\{ \ma{C}_{\ma p}\},
\end{equation}
where $\ma{C}_{\ma p}$ is the upper left $3 \times 3$ block of $\ma{J}^{-1}$. Using the blocks defined in \eqref{eq:block_fim}, $\ma{C}_{\ma p}$ can be straightforwardly computed using the Schur complement as
\begin{align*}
    \ma{Q} = \ma{J}_{a,\varphi} - \ma{J}_{{\ma p}, (a, \varphi)}^\trans \ma{J}_p^{-1} \ma{J}_{{\ma p}, (a, \varphi)} \\
    \ma{C}_{\ma p} = \frac{\sigma_n^2}{2} \left( \ma{J}_{\ma p}^{-1} + \ma{J}_{\ma p}^{-1} \ma{J}_{p, (a, \varphi)} \ma{Q}^{-1} \ma{J}_{{\ma p}, (a, \varphi)}^\trans \ma{J}_{\ma p}^{-1} \right).
\end{align*}
A direct observation from \eqref{eq:fim} is that if the total number of samples given by $\ma{\Phi}$ is too small, the \gls{fim} becomes singular. However, we can distribute the minimum number of Fourier coefficients over the complete set of spatial scan positions, which is usually much larger. Further, neighboring A-scans are highly correlated, since the physical phenomena are rather smooth on this scale. Due to this, measurements at position $(x,y)$ also yield a certain amount of information about $(x \pm \Delta x,y \pm \Delta y)$ such that choosing different mixing patterns $\bm \sigma_{n_x,n_y}$ or subselections $\bm S_{n_x,n_y}$ yields more information about the specimen.
Fig.~\ref{fig:crb_results} shows $\ma{c}$ for varying $n_f$, normalized by the \gls{crb} obtained by sampling the full spectrum, $C_{\rm N_t, \cdot}$. Since subsampling can only increase the \gls{crb}, we have that $C_{\rm N_t, \cdot} \leq C_{\cdot}$. Simulations are performed using the parameters in Tab.~\ref{tab:simulation_parameters} placing a single point source at a depth of~\SI{33.3}{mm} beneath the center of the scanning grid. The randomized strategies were averaged over 50 trials. The transparent areas represent the range between the best and worst \gls{crb} for the randomized strategies. From the figure, the following can be noted. For larger $n_f$, the knowledge based sampling performs better than using $\ma{\Phi}_{\rm rnd}$, since the bandwidth of the pulse is used up completely after a certain value for $n_f$. Further, $\ma{\Phi}_{\rm max}$ provides a lower bound to $\ma{\Phi}_{\rm nrg}$ in the considered scenario. This is due to the assumed Gaussian pulse shape, which leads to sampling more energy being better than sampling a higher bandwidth. For small $n_f$, the loss in performance is the highest for $\sqrt{C}_z$, which is expectable since the compression along individual A-scans deteriorates the delay estimation of the echoes the most. This in turn influences the estimation of the scatterers' depths the most. Varying the coefficients for each scan position leads to a significant performance gain when using $\ma{\Phi}_{{\rm rnd}, f}$ or $\ma{\Phi}_{{\rm nrg}, f}$ and reduces the overall loss compared to  $C_{\rm N_t, \cdot}$ to a factor of 2-3. 
%

\section{Numerical Simulations}\label{sec:num_sims}
In the following, we will present several reconstructions using \gls{fista}. The regularization parameter $\lambda$ is chosen as
\[
    \lambda = \mu \|\bm \beta_{\bm \Phi \bm H}(\ma{y})\|_\infty,
\]
with $0 < \mu < 1$.
\subsection{Experiments on simulated data\label{sec:simulated_data}}
We first consider reconstructions of simulated data sets to showcase the performance of the different sampling strategies, where simulate a shoe-box steel specimen. The parameters used for the simulation are listed in Tab.~\ref{tab:simulation_parameters} and the model for the pulse shape and transducer directivity are the same as in Sec.~\ref{sec:asymptotic_performance}. The imaging region starts at a depth of $z_d=\SI{29.6}{mm}$. All data sets presented in this section have been simulated noise free by simply evaluating $\ma{Ha}$ for a previously defined $\ma{a}$ as the goal is to compare only the performance of the different sampling strategies. The ground truth $\ma{a}$ is designed by setting $[\ma{a}]_d ={\rm e}^{\jmath \pi/4}, d \in \mathcal{D}$ and zero otherwise, where $\mathcal{D}$ is the set of indices forming the defect as a sum of point sources. The amplitude is chosen since it represents the largest possible phase offset that can arise from a scatterer actually lying ``between'' two grid points. It is therefore also the most challenging scenario for the popular choice of $\textit{Re}\lbrace \ma{H} \rbrace$ as a forward model. As targets we consider two types of defects. In the first scenario, the specimen contains a straight line simulating a defect at depth $25 z/\Delta z$. In the second scenario, the specimen contains a rectangle of size $40 \sqrt{(z/\Delta z)^2 + (x/\Delta x)^2} \times 40 y/\Delta y$ diagonally placed along the $z$-$x$-plane as a defect. The resulting datasets can be compressed by applying different incarnations of $\ma{\Phi}$ to $\ma{Ha}$. The compressed datasets are then reconstructed using \gls{fista} and the largest singular value is computed using \gls{arpack} if not explicitly stated otherwise. The results are plotted using a so-called C-scan image, i.e. we plot the maximum amplitude of an axis along the two remaining axes.\par
\begin{table}
    \begin{center}
    \begin{tabular}{l|c*{3}{|c}|c}
        $N_x$ & $N_y$ & $N_z$  & $\Delta x$ & $\Delta y$ & $c_0$   \\
        \hline
        50 & 50 & \SI{20}{\mega \hertz} & \SI{0.5}{mm} & \SI{0.5}{mm} & \SI{5920}{\meter \per \second}  \\
        \hline
        $\theta$ & $N_z$ & $f_c$ & $t_0$ & $z_d$ & $\alpha$ \\
        \hline
        $30^\circ$ & 50 & \SI{3.2}{\mega\hertz} &  \SI{10}{\micro \second} & \SI{29.6}{\milli \meter}  & $(0.65 f_c)^2$
    \end{tabular}
    \end{center}
    \caption{Simulation parameters \label{tab:simulation_parameters}}
\end{table}
As a first scenario, we compare the reconstructions using different subsampling strategies sampling only $n_f=1$ coefficient per A-scan. In addition to the different sampling strategies, different versions of the model matrix $\ma{H}$ can be used for reconstruction: (i) $\ma{H}$ represents the complex analytic model as defined in Sec.~\ref{sec:data_model}, (ii) $\text{Re}\{\ma{H}\}$ is equivalent to modeling $h(t)$ as a Gaussian windowed cosine, cf.~\cite{semper_eusipco_2019,laroche_ieeeuffc_2020}. The latter is only able to reconstruct a real-valued version of $\ma{a}$. Further, with respect to (ii), there exist two approaches on how to define $\ma{S}_i$ for the reconstruction: $\ma{\Phi}_{{\rm max}} \text{Re}\{\ma{H}\}$ inputs only the $n_f$ coefficients at one half of the symmetric spectrum into the reconstruction algorithm, ``$\ma{\Phi}_{{\rm max}} \text{Re}\{\ma{H}\}$ mirrored'' mirrors these coefficients to the other half. By this, essentially $2 n_f$ coefficients are input but only $n_f$ coefficients need to be measured at each scan. The latter reproduces the setup as previously published in~\cite{semper_eusipco_2019}. \par
\begin{figure}
\begin{tikzpicture}

\begin{groupplot}[group style={group name=upper group, group size=2 by 2, vertical sep=0, horizontal sep=0.5cm}]
\nextgroupplot[
axis equal image,
axis on top=true,
height= ,
minor xtick={},
minor ytick={},
scaled x ticks=manual:{}{\pgfmathparse{#1}},
tick align=outside,
tick pos=left,
title={\(\displaystyle \ma{\Phi}_{{\rm rnd},m}\ma{H}\)},
width=0.6\linewidth,
x grid style={white!69.0196078431373!black},
xmin=15, xmax=35,
xtick style={color=black},
xtick={10,20,30,40},
xticklabels={},
y grid style={white!69.0196078431373!black},
ylabel={\(\displaystyle x/\Delta x\)},
ymin=15, ymax=35,
ytick style={color=black},
ytick={10,20,30,40},
yticklabels={\(\displaystyle 10\),\(\displaystyle 20\),\(\displaystyle 30\),\(\displaystyle 40\)}
]
\addplot graphics [includegraphics cmd=\pgfimage,xmin=15, xmax=35, ymin=15, ymax=35] {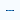};

\nextgroupplot[
axis equal image,
axis on top=true,
height= ,
minor xtick={},
minor ytick={},
scaled x ticks=manual:{}{\pgfmathparse{#1}},
scaled y ticks=manual:{}{\pgfmathparse{#1}},
tick align=outside,
tick pos=left,
title={\(\displaystyle \ma{\Phi}_{{\rm nrg},f} \text{Re}\{\ma{H}\}\)},
width=0.6\linewidth,
x grid style={white!69.0196078431373!black},
xmin=15, xmax=35,
xtick style={color=black},
xtick={10,20,30,40},
xticklabels={},
y grid style={white!69.0196078431373!black},
ymin=15, ymax=35,
ytick style={color=black},
ytick={10,20,30,40},
yticklabels={}
]
\addplot graphics [includegraphics cmd=\pgfimage,xmin=15, xmax=35, ymin=15, ymax=35] {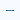};

\nextgroupplot[
axis equal image,
axis on top=true,
height= ,
minor xtick={},
minor ytick={},
scaled x ticks=manual:{}{\pgfmathparse{#1}},
tick align=outside,
tick pos=left,
width=0.6\linewidth,
x grid style={white!69.0196078431373!black},
xlabel={\(\displaystyle y/\Delta y\)},
xmin=15, xmax=35,
xtick style={color=black},
xtick={10,20,30,40},
xticklabels={},
y dir=reverse,
y grid style={white!69.0196078431373!black},
ylabel={\(\displaystyle z/\Delta z\)},
ymin=15, ymax=35,
ytick style={color=black},
ytick={15,25,35},
yticklabels={\(\displaystyle 15\),\(\displaystyle 25\),\(\displaystyle 35\)}
]
\addplot graphics [includegraphics cmd=\pgfimage,xmin=15, xmax=35, ymin=15, ymax=35] {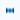};

\nextgroupplot[
axis equal image,
axis on top=true,
height= ,
minor xtick={},
minor ytick={},
scaled x ticks=manual:{}{\pgfmathparse{#1}},
scaled y ticks=manual:{}{\pgfmathparse{#1}},
tick align=outside,
tick pos=left,
width=0.6\linewidth,
x grid style={white!69.0196078431373!black},
xlabel={\(\displaystyle y/\Delta y\)},
xmin=15, xmax=35,
xtick style={color=black},
xtick={10,20,30,40},
xticklabels={},
y dir=reverse,
y grid style={white!69.0196078431373!black},
ymin=15, ymax=35,
ytick style={color=black},
ytick={15,25,35},
yticklabels={}
]
\addplot graphics [includegraphics cmd=\pgfimage,xmin=15, xmax=35, ymin=15, ymax=35] {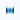};
\end{groupplot}
\node[above = 2cm of upper group c1r1.west, align=left] {(a)};
\end{tikzpicture}
\begin{tikzpicture}
\begin{groupplot}[group style={group name=lower group, group size=2 by 2, vertical sep=0, horizontal sep=0.5cm}]
\nextgroupplot[
axis equal image,
axis on top=true,
height= ,
minor xtick={},
minor ytick={},
scaled x ticks=manual:{}{\pgfmathparse{#1}},
tick align=outside,
tick pos=left,
title={\(\displaystyle \ma{\Phi}_{{\rm rnd},f}\ma{H}\)},
width=0.6\linewidth,
x grid style={white!69.0196078431373!black},
xmin=15, xmax=35,
xtick style={color=black},
xtick={10,20,30,40},
xticklabels={},
y grid style={white!69.0196078431373!black},
ylabel={\(\displaystyle x/\Delta x\)},
ymin=15, ymax=35,
ytick style={color=black},
ytick={10,20,30,40},
yticklabels={\(\displaystyle 10\),\(\displaystyle 20\),\(\displaystyle 30\),\(\displaystyle 40\)}
]
\addplot graphics [includegraphics cmd=\pgfimage,xmin=15, xmax=35, ymin=15, ymax=35] {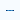};

\nextgroupplot[
axis equal image,
axis on top=true,
height= ,
minor xtick={},
minor ytick={},
scaled x ticks=manual:{}{\pgfmathparse{#1}},
scaled y ticks=manual:{}{\pgfmathparse{#1}},
tick align=outside,
tick pos=left,
title={\(\displaystyle \ma{\Phi}_{{\rm max}} \text{Re}\{\ma{H}\}\) mirrored},
width=0.6\linewidth,
x grid style={white!69.0196078431373!black},
xmin=15, xmax=35,
xtick style={color=black},
xtick={10,20,30,40},
xticklabels={},
y grid style={white!69.0196078431373!black},
ymin=15, ymax=35,
ytick style={color=black},
ytick={10,20,30,40},
yticklabels={}
]
\addplot graphics [includegraphics cmd=\pgfimage,xmin=15, xmax=35, ymin=15, ymax=35] {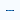};

\nextgroupplot[
axis equal image,
axis on top=true,
height= ,
minor xtick={},
minor ytick={},
tick align=outside,
tick pos=left,
width=0.6\linewidth,
x grid style={white!69.0196078431373!black},
xlabel={\(\displaystyle y/\Delta y\)},
xmin=15, xmax=35,
xtick style={color=black},
xtick={10,20,30,40},
xticklabels={\(\displaystyle 10\),\(\displaystyle 20\),\(\displaystyle 30\),\(\displaystyle 40\)},
y dir=reverse,
y grid style={white!69.0196078431373!black},
ylabel={\(\displaystyle z/\Delta z\)},
ymin=15, ymax=35,
ytick style={color=black},
ytick={15,25,35},
yticklabels={\(\displaystyle 15\),\(\displaystyle 25\),\(\displaystyle 35\)}
]
\addplot graphics [includegraphics cmd=\pgfimage,xmin=15, xmax=35, ymin=15, ymax=35] {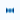};

\nextgroupplot[
axis equal image,
axis on top=true,
height= ,
minor xtick={},
minor ytick={},
scaled y ticks=manual:{}{\pgfmathparse{#1}},
tick align=outside,
tick pos=left,
width=0.6\linewidth,
x grid style={white!69.0196078431373!black},
xlabel={\(\displaystyle y/\Delta y\)},
xmin=15, xmax=35,
xtick style={color=black},
xtick={10,20,30,40},
xticklabels={\(\displaystyle 10\),\(\displaystyle 20\),\(\displaystyle 30\),\(\displaystyle 40\)},
y dir=reverse,
y grid style={white!69.0196078431373!black},
ymin=15, ymax=35,
ytick style={color=black},
ytick={15,25,35},
yticklabels={}
]
\addplot graphics [includegraphics cmd=\pgfimage,xmin=15, xmax=35, ymin=15, ymax=35] {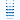};
\end{groupplot}
\node[above = 2cm of lower group c1r1.west, align=left] {(b)};
\end{tikzpicture}
    \caption{Reconstruction results from simulated data for $n_f=1$ using \gls{fista} with $\mu=0.6$ stopped after 80 iterations. \label{fig:simulated_comparison}}
\end{figure}

The results are depicted in Fig.~\ref{fig:simulated_comparison}.  All strategies correctly reconstruct the shape of the target in the $x$-$y$ plane since the lateral focusing is not deteriorated (which is also in accordance with the asymptotic results in Sec.~\ref{sec:asymptotic_performance}).
Note that the difference between the second column of Fig.~\ref{fig:simulated_comparison}~(a) and the second column of Fig.~\ref{fig:simulated_comparison}~(b) is that in (a) only one half of the spectrum is sampled although a real-valued matrix $\ma{H}$ is used.
Using a real-valued model as well as a constant $\ma{\Phi}$ for all scan positions results in the worst localization in depth. In addition, the peak along $z$ in the reconstruction is at a wrong position due to the defect being off-grid (phase offset) (cf. Fig.~\ref{fig:simulated_comparison}~(b), second column). In contrast, as expected from the \gls{crb}, using distinct subsampling patterns restores the localization accuracy along the $z$-axis. Further, it can be seen that in the case of random uniform subsampling, it is equivalent to vary the mixing pattern or the set of Fourier coefficients per A-scan. Note that $\ma{\Phi}_{{\rm rnd},f}\ma{H}$ and $\ma{\Phi}_{{\rm nrg},f} \ma{H}$ perform equivalently in this scenario, which is why the latter is not depicted. Finally, it can be noted that using $\ma{H}$ instead of $\text{Re}\{\ma{H}\}$ leads to a slightly sparser solution.\par

\begin{figure*}
    \begin{center}
\begin{tikzpicture}

\begin{groupplot}[group style={group size=4 by 3, vertical sep=0.7cm, horizontal sep=0.5cm}]
\nextgroupplot[
axis equal image,
axis on top=true,
height=0.25\textwidth,
minor xtick={},
minor ytick={},
scaled x ticks=manual:{}{\pgfmathparse{#1}},
tick align=outside,
tick pos=left,
title={\(\displaystyle n_f=1\)},
width=0.25\textwidth,
x grid style={white!69.0196078431373!black},
xmin=0, xmax=50,
xtick style={color=black},
xtick={0,25,50},
xticklabels={},
y dir=reverse,
y grid style={white!69.0196078431373!black},
ylabel={\(\displaystyle z/\Delta z\)},
ymin=0, ymax=50,
ytick style={color=black},
ytick={0,25,50}
]
\addplot graphics [includegraphics cmd=\pgfimage,xmin=-0.5, xmax=49.5, ymin=49.5, ymax=-0.5] {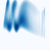};

\nextgroupplot[
axis equal image,
axis on top=true,
height=0.25\textwidth,
minor xtick={},
minor ytick={},
scaled x ticks=manual:{}{\pgfmathparse{#1}},
scaled y ticks=manual:{}{\pgfmathparse{#1}},
tick align=outside,
tick pos=left,
title={\(\displaystyle n_f=2\)},
width=0.25\textwidth,
x grid style={white!69.0196078431373!black},
xmin=0, xmax=50,
xtick style={color=black},
xtick={0,25,50},
xticklabels={},
y dir=reverse,
y grid style={white!69.0196078431373!black},
ymin=0, ymax=50,
ytick style={color=black},
ytick={0,25,50},
yticklabels={}
]
\addplot graphics [includegraphics cmd=\pgfimage,xmin=-0.5, xmax=49.5, ymin=49.5, ymax=-0.5] {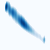};

\nextgroupplot[
axis equal image,
axis on top=true,
height=0.25\textwidth,
minor xtick={},
minor ytick={},
scaled x ticks=manual:{}{\pgfmathparse{#1}},
scaled y ticks=manual:{}{\pgfmathparse{#1}},
tick align=outside,
tick pos=left,
title={\(\displaystyle n_f=5\)},
width=0.25\textwidth,
x grid style={white!69.0196078431373!black},
xmin=0, xmax=50,
xtick style={color=black},
xtick={0,25,50},
xticklabels={},
y dir=reverse,
y grid style={white!69.0196078431373!black},
ymin=0, ymax=50,
ytick style={color=black},
ytick={0,25,50},
yticklabels={}
]
\addplot graphics [includegraphics cmd=\pgfimage,xmin=-0.5, xmax=49.5, ymin=49.5, ymax=-0.5] {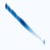};

\nextgroupplot[
axis equal image,
axis on top=true,
height=0.25\textwidth,
minor xtick={},
minor ytick={},
scaled x ticks=manual:{}{\pgfmathparse{#1}},
scaled y ticks=manual:{}{\pgfmathparse{#1}},
tick align=outside,
tick pos=left,
title={\(\displaystyle n_f=10\)},
width=0.25\textwidth,
x grid style={white!69.0196078431373!black},
xmin=0, xmax=50,
xtick style={color=black},
xtick={0,25,50},
xticklabels={},
y dir=reverse,
y grid style={white!69.0196078431373!black},
ymin=0, ymax=50,
ytick style={color=black},
ytick={0,25,50},
yticklabels={}
]
\addplot graphics [includegraphics cmd=\pgfimage,xmin=-0.5, xmax=49.5, ymin=49.5, ymax=-0.5] {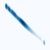};

\nextgroupplot[
axis equal image,
axis on top=true,
height=0.25\textwidth,
minor xtick={},
minor ytick={},
scaled x ticks=manual:{}{\pgfmathparse{#1}},
tick align=outside,
tick pos=left,
width=0.25\textwidth,
x grid style={white!69.0196078431373!black},
xmin=0, xmax=50,
xtick style={color=black},
xtick={0,25,50},
xticklabels={},
y dir=reverse,
y grid style={white!69.0196078431373!black},
ylabel={\(\displaystyle z/\Delta z\)},
ymin=0, ymax=50,
ytick style={color=black},
ytick={0,25,50}
]
\addplot graphics [includegraphics cmd=\pgfimage,xmin=-0.5, xmax=49.5, ymin=49.5, ymax=-0.5] {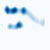};

\nextgroupplot[
axis equal image,
axis on top=true,
height=0.25\textwidth,
minor xtick={},
minor ytick={},
scaled x ticks=manual:{}{\pgfmathparse{#1}},
scaled y ticks=manual:{}{\pgfmathparse{#1}},
tick align=outside,
tick pos=left,
width=0.25\textwidth,
x grid style={white!69.0196078431373!black},
xmin=0, xmax=50,
xtick style={color=black},
xtick={0,25,50},
xticklabels={},
y dir=reverse,
y grid style={white!69.0196078431373!black},
ymin=0, ymax=50,
ytick style={color=black},
ytick={0,25,50},
yticklabels={}
]
\addplot graphics [includegraphics cmd=\pgfimage,xmin=-0.5, xmax=49.5, ymin=49.5, ymax=-0.5] {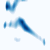};

\nextgroupplot[
axis equal image,
axis on top=true,
height=0.25\textwidth,
minor xtick={},
minor ytick={},
scaled x ticks=manual:{}{\pgfmathparse{#1}},
scaled y ticks=manual:{}{\pgfmathparse{#1}},
tick align=outside,
tick pos=left,
width=0.25\textwidth,
x grid style={white!69.0196078431373!black},
xmin=0, xmax=50,
xtick style={color=black},
xtick={0,25,50},
xticklabels={},
y dir=reverse,
y grid style={white!69.0196078431373!black},
ymin=0, ymax=50,
ytick style={color=black},
ytick={0,25,50},
yticklabels={}
]
\addplot graphics [includegraphics cmd=\pgfimage,xmin=-0.5, xmax=49.5, ymin=49.5, ymax=-0.5] {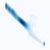};

\nextgroupplot[
axis equal image,
axis on top=true,
height=0.25\textwidth,
minor xtick={},
minor ytick={},
scaled x ticks=manual:{}{\pgfmathparse{#1}},
scaled y ticks=manual:{}{\pgfmathparse{#1}},
tick align=outside,
tick pos=left,
width=0.25\textwidth,
x grid style={white!69.0196078431373!black},
xmin=0, xmax=50,
xtick style={color=black},
xtick={0,25,50},
xticklabels={},
y dir=reverse,
y grid style={white!69.0196078431373!black},
ymin=0, ymax=50,
ytick style={color=black},
ytick={0,25,50},
yticklabels={}
]
\addplot graphics [includegraphics cmd=\pgfimage,xmin=-0.5, xmax=49.5, ymin=49.5, ymax=-0.5] {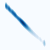};

\nextgroupplot[
axis equal image,
axis on top=true,
height=0.25\textwidth,
minor xtick={},
minor ytick={},
tick align=outside,
tick pos=left,
width=0.25\textwidth,
x grid style={white!69.0196078431373!black},
xlabel={\(\displaystyle x/\Delta x\)},
xmin=0, xmax=50,
xtick style={color=black},
xtick={0,25,50},
y dir=reverse,
y grid style={white!69.0196078431373!black},
ylabel={\(\displaystyle z/\Delta z\)},
ymin=0, ymax=50,
ytick style={color=black},
ytick={0,25,50}
]
\addplot graphics [includegraphics cmd=\pgfimage,xmin=-0.5, xmax=49.5, ymin=49.5, ymax=-0.5] {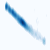};

\nextgroupplot[
axis equal image,
axis on top=true,
height=0.25\textwidth,
minor xtick={},
minor ytick={},
scaled y ticks=manual:{}{\pgfmathparse{#1}},
tick align=outside,
tick pos=left,
width=0.25\textwidth,
x grid style={white!69.0196078431373!black},
xlabel={\(\displaystyle x/\Delta x\)},
xmin=0, xmax=50,
xtick style={color=black},
xtick={0,25,50},
y dir=reverse,
y grid style={white!69.0196078431373!black},
ymin=0, ymax=50,
ytick style={color=black},
ytick={0,25,50},
yticklabels={}
]
\addplot graphics [includegraphics cmd=\pgfimage,xmin=-0.5, xmax=49.5, ymin=49.5, ymax=-0.5] {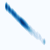};

\nextgroupplot[
axis equal image,
axis on top=true,
height=0.25\textwidth,
minor xtick={},
minor ytick={},
scaled y ticks=manual:{}{\pgfmathparse{#1}},
tick align=outside,
tick pos=left,
width=0.25\textwidth,
x grid style={white!69.0196078431373!black},
xlabel={\(\displaystyle x/\Delta x\)},
xmin=0, xmax=50,
xtick style={color=black},
xtick={0,25,50},
y dir=reverse,
y grid style={white!69.0196078431373!black},
ymin=0, ymax=50,
ytick style={color=black},
ytick={0,25,50},
yticklabels={}
]
\addplot graphics [includegraphics cmd=\pgfimage,xmin=-0.5, xmax=49.5, ymin=49.5, ymax=-0.5] {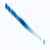};

\nextgroupplot[
axis equal image,
axis on top=true,
height=0.25\textwidth,
minor xtick={},
minor ytick={},
scaled y ticks=manual:{}{\pgfmathparse{#1}},
tick align=outside,
tick pos=left,
title={Ground Truth},
title style={yshift=-1.5ex},
width=0.25\textwidth,
x grid style={white!69.0196078431373!black},
xlabel={\(\displaystyle x/\Delta x\)},
xmin=0, xmax=50,
xtick style={color=black},
xtick={0,25,50},
y dir=reverse,
y grid style={white!69.0196078431373!black},
ymin=0, ymax=50,
ytick style={color=black},
ytick={0,25,50},
yticklabels={}
]
\addplot graphics [includegraphics cmd=\pgfimage,xmin=-0.5, xmax=49.5, ymin=49.5, ymax=-0.5] {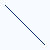};
\end{groupplot}

\end{tikzpicture}\end{center}
    \caption{Left to right: $n_f=1,2,5,10$, except for the figure entitled Ground Truth. Top to bottom: $\ma{\Phi}_{\rm max}$, $\ma{\Phi}_{\rm rnd}$, $\ma{\Phi}_{{\rm nrg}, f}$. Note that the missing result for $\ma{\Phi}_{{\rm nrg}, f}$ with $n_f = 10$ is visually equivalent to the reconstruction for $\ma{\Phi}_{{\rm nrg}, f}$ with $n_f=5$. Reconstructed using \gls{fista} with $\mu=0.1$, stopped after 80 iterations.\label{fig:simulations_square}}
\end{figure*}

As another scenario, we compare the subsampling schemes for varying $n_f$ in a more complex scenario. The results are depicted in Fig.~\ref{fig:simulations_square} (the bottom right figure shows the side view of the square as ground truth). The chosen scenario results in a measurement, where every A-scan sees echoes from almost all point sources. The ramifications of this are as follows: Choosing a constant subsampling pattern for all A-scans leads to a failed reconstruction for $n_f$ too small (first and second row.). This is basically also what our findings about the \gls{crb} already infer analytically. On the other hand, varying the pattern (third row) strongly improves the reconstructed image. This again emphasizes that the large number of (necessary) spatial scanning positions reduces the number of required temporal measurements. The spatial grid spacing needs to be small enough to ensure that defects with a given minimum size are still detected. However, due to this small spatial grid spacing, the scan at position $(x,y)$ also yields a certain amount of information about $(x \pm \Delta x,y \pm \Delta y)$.
\subsection{Synthetic aperture measurements \label{sec:synthetic_aperture_measurements}}
In the following, we compare the different approaches defined in Sec.~\ref{sec:fourier_subsampling} based on synthetic aperture measurements of a steel specimen. The specimen contains several flat bottom holes with diameters $\varnothing \SI{2}{\milli \meter}, \ \varnothing \SI{3}{\milli \meter}, \ \text{and} \ \varnothing \SI{5}{\milli \meter}$, representing artificial flaws. The measurements were taken using time domain sampling at a sampling rate of $f_s = \SI{20}{\mega\hertz}$. Fourier subsampling was simulated by calculating an \gls{fft} and keeping only $n_f$ Fourier coefficients per A-scan based on the chosen strategy. To calculate $\ma{H}$ the same Gaussian pulse model and transducer directivity as above is used. The center frequency of the transducer is at $\omega_c = 2\pi f_c$ and $\alpha =  (0.8 f_c)^2$. The specimen was scanned with $\Delta x = \Delta y = \SI{0.5}{\milli \meter}$. The opening angle of the transducer is set to $\theta = 30^\circ$.  The speed of sound in this steel is assumed to be $c_0 = \SI{5920}{\meter \per \second}$. For the reconstruction, \gls{fista} is used with $\mu = 0.4$ and stopped after 20 iterations. \par

\begin{figure*}
\begin{tikzpicture}

\begin{groupplot}[group style={group size=1 by 2, vertical sep=0, x descriptions at=edge bottom}]
\nextgroupplot[
axis equal image,
axis on top=true,
height= ,
scaled x ticks=manual:{}{\pgfmathparse{#1}},
tick align=outside,
tick pos=left,
width=\textwidth,
x grid style={white!69.0196078431373!black},
xmin=0, xmax=190.5,
xtick style={color=black},
xticklabels={},
y grid style={white!69.0196078431373!black},
ylabel={\(\displaystyle y\) in mm},
ymin=0, ymax=35,
ytick style={color=black},
ytick={0,20,40},
yticklabels={\(\displaystyle 0\),\(\displaystyle 20\),\(\displaystyle 40\)}
]
\addplot graphics [includegraphics cmd=\pgfimage,xmin=0, xmax=190.5, ymin=0, ymax=35] {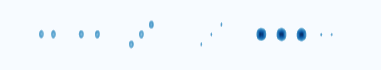};
\input{figures/muse_ground_truth_top_view.tex}

\nextgroupplot[
axis equal image,
axis on top=true,
height= ,
tick align=outside,
tick pos=left,
width=\textwidth,
x grid style={white!69.0196078431373!black},
xlabel={\(\displaystyle x\) in mm},
xmin=0, xmax=190.5,
xtick style={color=black},
xtick={0,25,50,75,100,125,150,175,200},
xticklabels={\(\displaystyle 0\),\(\displaystyle 25\),\(\displaystyle 50\),\(\displaystyle 75\),\(\displaystyle 100\),\(\displaystyle 125\),\(\displaystyle 150\),\(\displaystyle 175\),\(\displaystyle 200\)},
y dir=reverse,
y grid style={white!69.0196078431373!black},
ylabel={\(\displaystyle z\) in mm},
ymin=67.1125, ymax=89.2375,
ytick style={color=black},
ytick={70,80},
yticklabels={\(\displaystyle 70\),\(\displaystyle 80\)}
]
\addplot graphics [includegraphics cmd=\pgfimage,xmin=0, xmax=190.5, ymin=67.1125, ymax=89.2375] {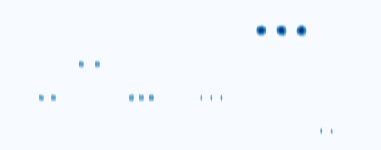};
\newcommand{\yPosOne}{18}
\newcommand{\yPosTwo}{13}
\newcommand{\circleRadiusOne}{1.5}
\newcommand{\circleRadiusTwo}{0.75}
\newcommand{\circleRadiusThree}{2.75}
\newcommand{\offsetOne}{20.7}
\newcommand{\offsetTwo}{65.7}
\newcommand{\offsetThree}{100.7}
\newcommand{\offsetFour}{160.7}
\newcommand{\offsetFive}{130.7}
\newcommand{\TenMMDeepHoles}{81.5}
\newcommand{\FifteenMMDeepHoles}{76.5}
\newcommand{\TwentyMMDeepHoles}{71.5}
\newcommand{\FiveMMDeepHoles}{86.5}
\newcommand{\linestyleGT}{densely dotted}
\newcommand{\linestyleFBHWalls}{dashed}

\draw[\linestyleGT] (\offsetOne-\circleRadiusOne, \TenMMDeepHoles) -- (\offsetOne+\circleRadiusOne, \TenMMDeepHoles) {};
\draw[\linestyleFBHWalls] (\offsetOne-\circleRadiusOne, 90) -- (\offsetOne-\circleRadiusOne, \TenMMDeepHoles){};
\draw[\linestyleFBHWalls] (\offsetOne+\circleRadiusOne, 90) -- (\offsetOne+\circleRadiusOne, \TenMMDeepHoles){};
\draw[\linestyleGT] (\offsetOne+6-\circleRadiusOne, \TenMMDeepHoles) -- (\offsetOne+6+\circleRadiusOne, \TenMMDeepHoles) {};
\draw[\linestyleFBHWalls] (\offsetOne+6-\circleRadiusOne, 90) -- (\offsetOne+6-\circleRadiusOne, \TenMMDeepHoles){};
\draw[\linestyleFBHWalls] (\offsetOne+6+\circleRadiusOne, 90) -- (\offsetOne+6+\circleRadiusOne, \TenMMDeepHoles){};
\draw[\linestyleGT] (\offsetOne+20-\circleRadiusOne, \FifteenMMDeepHoles) -- (\offsetOne+20+\circleRadiusOne, \FifteenMMDeepHoles) {};
\draw[\linestyleFBHWalls] (\offsetOne+20-\circleRadiusOne, 90) -- (\offsetOne+20-\circleRadiusOne, \FifteenMMDeepHoles){};
\draw[\linestyleFBHWalls] (\offsetOne+20+\circleRadiusOne, 90) -- (\offsetOne+20+\circleRadiusOne, \FifteenMMDeepHoles){};
\draw[\linestyleGT] (\offsetOne+28-\circleRadiusOne, \FifteenMMDeepHoles) -- (\offsetOne+28+\circleRadiusOne, \FifteenMMDeepHoles) {};
\draw[\linestyleFBHWalls] (\offsetOne+28-\circleRadiusOne, 90) -- (\offsetOne+28-\circleRadiusOne, \FifteenMMDeepHoles){};
\draw[\linestyleFBHWalls] (\offsetOne+28+\circleRadiusOne, 90) -- (\offsetOne+28+\circleRadiusOne, \FifteenMMDeepHoles){};

\draw[\linestyleGT] (\offsetTwo-\circleRadiusOne, \TenMMDeepHoles) -- (\offsetTwo+\circleRadiusOne, \TenMMDeepHoles) {};
\draw[\linestyleFBHWalls] (\offsetTwo-\circleRadiusOne, 90) -- (\offsetTwo-\circleRadiusOne, \TenMMDeepHoles) {};
\draw[\linestyleFBHWalls] (\offsetTwo+\circleRadiusOne, 90) -- (\offsetTwo+\circleRadiusOne, \TenMMDeepHoles) {};
\draw[\linestyleGT] (\offsetTwo+5-\circleRadiusOne, \TenMMDeepHoles) -- (\offsetTwo+5+\circleRadiusOne, \TenMMDeepHoles) {};
\draw[\linestyleFBHWalls] (\offsetTwo+5-\circleRadiusOne, 90) -- (\offsetTwo+5-\circleRadiusOne, \TenMMDeepHoles) {};
\draw[\linestyleFBHWalls] (\offsetTwo+5+\circleRadiusOne, 90) -- (\offsetTwo+5+\circleRadiusOne, \TenMMDeepHoles) {};
\draw[\linestyleGT] (\offsetTwo+10-\circleRadiusOne, \TenMMDeepHoles) -- (\offsetTwo+10+\circleRadiusOne, \TenMMDeepHoles) {};
\draw[\linestyleFBHWalls] (\offsetTwo+10-\circleRadiusOne, 90) -- (\offsetTwo+10-\circleRadiusOne, \TenMMDeepHoles) {};
\draw[\linestyleFBHWalls] (\offsetTwo+10+\circleRadiusOne, 90) -- (\offsetTwo+10+\circleRadiusOne, \TenMMDeepHoles) {};

\draw[\linestyleGT] (\offsetThree-\circleRadiusTwo, \TenMMDeepHoles) -- (\offsetThree+\circleRadiusTwo, \TenMMDeepHoles) {};
\draw[\linestyleFBHWalls] (\offsetThree-\circleRadiusTwo, 90) -- (\offsetThree-\circleRadiusTwo, \TenMMDeepHoles) {};
\draw[\linestyleFBHWalls] (\offsetThree+\circleRadiusTwo, 90) -- (\offsetThree+\circleRadiusTwo, \TenMMDeepHoles) {};
\draw[\linestyleGT] (\offsetThree+5-\circleRadiusTwo, \TenMMDeepHoles) -- (\offsetThree+5+\circleRadiusTwo, \TenMMDeepHoles) {};
\draw[\linestyleFBHWalls] (\offsetThree+5-\circleRadiusTwo, 90) -- (\offsetThree+5-\circleRadiusTwo, \TenMMDeepHoles) {};
\draw[\linestyleFBHWalls] (\offsetThree+5+\circleRadiusTwo, 90) -- (\offsetThree+5+\circleRadiusTwo, \TenMMDeepHoles) {};
\draw[\linestyleGT] (\offsetThree+10-\circleRadiusTwo, \TenMMDeepHoles) -- (\offsetThree+10+\circleRadiusTwo, \TenMMDeepHoles) {};
\draw[\linestyleFBHWalls] (\offsetThree+10-\circleRadiusTwo, 90) -- (\offsetThree+10-\circleRadiusTwo, \TenMMDeepHoles) {};
\draw[\linestyleFBHWalls] (\offsetThree+10+\circleRadiusTwo, 90) -- (\offsetThree+10+\circleRadiusTwo, \TenMMDeepHoles) {};

\draw[\linestyleGT] (\offsetFive-\circleRadiusThree, \TwentyMMDeepHoles) -- (\offsetFive+\circleRadiusThree, \TwentyMMDeepHoles) {};
\draw[\linestyleFBHWalls] (\offsetFive-\circleRadiusThree, 90) --
(\offsetFive-\circleRadiusThree, \TwentyMMDeepHoles) {};
\draw[\linestyleFBHWalls] (\offsetFive+\circleRadiusThree, 90) --
(\offsetFive+\circleRadiusThree, \TwentyMMDeepHoles) {};
\draw[\linestyleGT] (\offsetFive+10-\circleRadiusThree, \TwentyMMDeepHoles) -- (\offsetFive+10+\circleRadiusThree, \TwentyMMDeepHoles) {};
\draw[\linestyleFBHWalls] (\offsetFive+10-\circleRadiusThree, 90) --
(\offsetFive+10-\circleRadiusThree, \TwentyMMDeepHoles) {};
\draw[\linestyleFBHWalls] (\offsetFive+10+\circleRadiusThree, 90) --
(\offsetFive+10+\circleRadiusThree, \TwentyMMDeepHoles) {};

\draw[\linestyleGT] (\offsetFive+20-\circleRadiusThree, \TwentyMMDeepHoles) -- (\offsetFive+20+\circleRadiusThree, \TwentyMMDeepHoles) {};
\draw[\linestyleFBHWalls] (\offsetFive+20-\circleRadiusThree, 90) --
(\offsetFive+20-\circleRadiusThree, \TwentyMMDeepHoles) {};
\draw[\linestyleFBHWalls] (\offsetFive+20+\circleRadiusThree, 90) --
(\offsetFive+20+\circleRadiusThree, \TwentyMMDeepHoles) {};

\draw[\linestyleGT] (\offsetFour-\circleRadiusTwo, \FiveMMDeepHoles) -- (\offsetFour+\circleRadiusTwo, \FiveMMDeepHoles) {};
\draw[\linestyleFBHWalls] (\offsetFour-\circleRadiusTwo,90) --(\offsetFour-\circleRadiusTwo, \FiveMMDeepHoles) {};
\draw[\linestyleFBHWalls] (\offsetFour+\circleRadiusTwo,90) --(\offsetFour+\circleRadiusTwo, \FiveMMDeepHoles) {};

\draw[\linestyleGT] (\offsetFour+5.25-\circleRadiusTwo, \FiveMMDeepHoles) -- (\offsetFour+5.25+\circleRadiusTwo, \FiveMMDeepHoles) {};
\draw[\linestyleFBHWalls] (\offsetFour+5.25-\circleRadiusTwo,90) --(\offsetFour+5.25-\circleRadiusTwo, \FiveMMDeepHoles) {};
\draw[\linestyleFBHWalls] (\offsetFour+5.25+\circleRadiusTwo,90) --(\offsetFour+5.25+\circleRadiusTwo, \FiveMMDeepHoles) {};
\end{groupplot}

\end{tikzpicture}
    \caption{Top and side view of a FISTA reconstruction using $n_f=1$ and $\ma{\Phi}_{{\rm rnd}, m,f} \ma{H}$. FISTA was run for 20 steps with $\mu=0.4$. \label{fig:muse_random_best}}
\end{figure*}

Fig.~\ref{fig:muse_random_best} shows the result using $\ma{\Phi}_{{\rm rnd}, m,f}$ using $n_f = 1$ per A-scan. The top figure shows a C-scan view from the top. The bottom image shows the projection onto the $y$-axis. In both cases a ground truth sketch is superimposed onto the image. It clearly shows that the reconstruction reproduces the positions of the defects accurately in all three dimensions as well as their extent in $x$-$y$-direction. \par
\begin{figure*}
\newcommand{\scalingFactor}{0.27}

\begin{tikzpicture}

\begin{groupplot}[group style={group size=8 by 1}]
\nextgroupplot[
axis equal image,
axis on top=true,
height= ,
minor xtick={},
minor ytick={},
tick align=outside,
tick pos=left,
width=\scalingFactor\textwidth,
x grid style={white!69.0196078431373!black},
xlabel={mm},
xmin=0, xmax=20,
xtick style={color=black},
xtick={0,10,20},
y grid style={white!69.0196078431373!black},
ylabel={mm},
ymin=0, ymax=42.5,
ytick style={color=black},
ytick={0,20,40}
]
\input{figures/ground_truth_muse_small.tex}

\nextgroupplot[
axis equal image,
axis on top=true,
height= ,
minor xtick={},
minor ytick={},
scaled y ticks=manual:{}{\pgfmathparse{#1}},
tick align=outside,
tick pos=left,
title={\(\displaystyle \ma{H}\)},
width=\scalingFactor\textwidth,
x grid style={white!69.0196078431373!black},
xlabel={mm},
xmin=0, xmax=20,
xtick style={color=black},
xtick={0,10,20},
y grid style={white!69.0196078431373!black},
ymin=0, ymax=42.5,
ytick style={color=black},
ytick={0,20,40},
yticklabels={}
]
\addplot graphics [includegraphics cmd=\pgfimage,xmin=0, xmax=20, ymin=0, ymax=42.5] {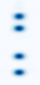};

\nextgroupplot[
axis equal image,
axis on top=true,
height= ,
minor xtick={},
minor ytick={},
scaled y ticks=manual:{}{\pgfmathparse{#1}},
tick align=outside,
tick pos=left,
title={\(\displaystyle \ma{\Phi}_{\rm max} \ma{H}\)},
width=\scalingFactor\textwidth,
x grid style={white!69.0196078431373!black},
xlabel={mm},
xmin=0, xmax=20,
xtick style={color=black},
xtick={0,10,20},
y grid style={white!69.0196078431373!black},
ymin=0, ymax=42.5,
ytick style={color=black},
ytick={0,20,40},
yticklabels={}
]
\addplot graphics [includegraphics cmd=\pgfimage,xmin=0, xmax=20, ymin=0, ymax=42.5] {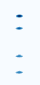};

\nextgroupplot[
axis equal image,
axis on top=true,
height= ,
minor xtick={},
minor ytick={},
scaled y ticks=manual:{}{\pgfmathparse{#1}},
tick align=outside,
tick pos=left,
title={\(\displaystyle \ma{\Phi}_{\rm rnd}\ma{H}\)},
width=\scalingFactor\textwidth,
x grid style={white!69.0196078431373!black},
xlabel={mm},
xmin=0, xmax=20,
xtick style={color=black},
xtick={0,10,20},
y grid style={white!69.0196078431373!black},
ymin=0, ymax=42.5,
ytick style={color=black},
ytick={0,20,40},
yticklabels={}
]
\addplot graphics [includegraphics cmd=\pgfimage,xmin=0, xmax=20, ymin=0, ymax=42.5] {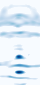};

\nextgroupplot[
axis equal image,
axis on top=true,
height= ,
minor xtick={},
minor ytick={},
scaled y ticks=manual:{}{\pgfmathparse{#1}},
tick align=outside,
tick pos=left,
title={\(\displaystyle \ma{\Phi}_{\rm nrg} \ma{H}\)},
width=\scalingFactor\textwidth,
x grid style={white!69.0196078431373!black},
xlabel={mm},
xmin=0, xmax=20,
xtick style={color=black},
xtick={0,10,20},
y grid style={white!69.0196078431373!black},
ymin=0, ymax=42.5,
ytick style={color=black},
ytick={0,20,40},
yticklabels={}
]
\addplot graphics [includegraphics cmd=\pgfimage,xmin=0, xmax=20, ymin=0, ymax=42.5] {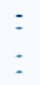};

\nextgroupplot[
axis equal image,
axis on top=true,
height= ,
minor xtick={},
minor ytick={},
scaled y ticks=manual:{}{\pgfmathparse{#1}},
tick align=outside,
tick pos=left,
title={\(\displaystyle \ma{\Phi}_{{\rm rnd},m} \ma{H}\)},
width=\scalingFactor\textwidth,
x grid style={white!69.0196078431373!black},
xlabel={mm},
xmin=0, xmax=20,
xtick style={color=black},
xtick={0,10,20},
y grid style={white!69.0196078431373!black},
ymin=0, ymax=42.5,
ytick style={color=black},
ytick={0,20,40},
yticklabels={}
]
\addplot graphics [includegraphics cmd=\pgfimage,xmin=0, xmax=20, ymin=0, ymax=42.5] {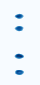};

\nextgroupplot[
axis equal image,
axis on top=true,
height= ,
minor xtick={},
minor ytick={},
scaled y ticks=manual:{}{\pgfmathparse{#1}},
tick align=outside,
tick pos=left,
title={\(\displaystyle \ma{\Phi}_{{\rm nrg}, f} \ma{H}\)},
width=\scalingFactor\textwidth,
x grid style={white!69.0196078431373!black},
xlabel={mm},
xmin=0, xmax=20,
xtick style={color=black},
xtick={0,10,20},
y grid style={white!69.0196078431373!black},
ymin=0, ymax=42.5,
ytick style={color=black},
ytick={0,20,40},
yticklabels={}
]
\addplot graphics [includegraphics cmd=\pgfimage,xmin=0, xmax=20, ymin=0, ymax=42.5] {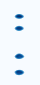};

\nextgroupplot[
axis equal image,
axis on top=true,
height= ,
minor xtick={},
minor ytick={},
scaled y ticks=manual:{}{\pgfmathparse{#1}},
tick align=outside,
tick pos=left,
title={\(\displaystyle \ma{\Phi}_{{\rm rnd}, m,f} \ma{H}\)},
width=\scalingFactor\textwidth,
x grid style={white!69.0196078431373!black},
xlabel={mm},
xmin=0, xmax=20,
xtick style={color=black},
xtick={0,10,20},
y grid style={white!69.0196078431373!black},
ymin=0, ymax=42.5,
ytick style={color=black},
ytick={0,20,40},
yticklabels={}
]
\addplot graphics [includegraphics cmd=\pgfimage,xmin=0, xmax=20, ymin=0, ymax=42.5] {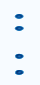};
\end{groupplot}

\end{tikzpicture}
    \caption{Zoomed comparison of the different approaches (top view). The second figure from the left shows the standard uncompressed \gls{saft} reconstruction. The six remaining figures show the reconstructions based on the different sampling schemes for $n_f=1$ . The reconstructions are computed using FISTA with $\mu=0.4$, stopped after 20 iterations. \label{fig:muse_zoom}}
\end{figure*}

Fig.~\ref{fig:muse_zoom} shows a zoomed in comparison of all investigated subsampling strategies as well as the conventional uncompressed \gls{saft} reconstruction. Comparing the top row with the bottom row reveals the advantage of varying the mixing pattern and/or the measured Fourier coefficient.
\begin{discussion}
\begin{itemize}[wide=0.5em, leftmargin =*, nosep, before = \leavevmode\vspace{-\baselineskip}]
    \item  Varying the sample patterns for each A-scan provides a significant gain compared to only using a single constant pattern. With this improvement, $n_f=1$ can already be enough to obtain high resolution reconstructions.
    \item $\text{Re}\{\ma{H}\}$ performs worse since it is incapable of dealing with off-grid contributions.
    \item The inclusion of the symmetric counterparts of the Fourier coefficients of the (real-valued) measurements into the reconstruction leads to a deterioration of the reconstruction.
    \item The \gls{crb} is a valid proxy to evaluate the performance of the different subsampling strategies. The improvement when varying the sampling or mixing patterns is well predicted.
    \item Both the \gls{crb} and the reconstruction show how the subsampling mainly affects the performance along the depth axis: the depth resolution can be abysmal while still maintaining a high quality C-scan image along the $x$-$y$ plane.
\end{itemize}
\end{discussion}


\section{Conclusion \label{sec:Conclusion}}
In this paper, we developed and analyzed a novel \gls{cs}-based ultrasound acquisition framework for synthetic aperture \gls{ndt}. 
The proposed new strategies $\ma{\Phi}_{\dots, f}$ and $\ma{\Phi}_{\dots, m}$ are superior to existing state of the art strategies. However, this is only true, if we consider a forward model that exploits the high correlation between adjacent scans such as the employed 3-D model for the reconstruction. By doing so, the number of necessary Fourier coefficients per scan can even be reduced to the absolute minimum of a single coefficient even in realistic scenarios.
In total, $\ma{\Phi}_{{\rm nrg}, f}$ provides the best choice, since it provides the best imaging performance (together with $\ma{\Phi}_{{\rm rnd}, m}$ and $\ma{\Phi}_{{\rm rnd}, f}$) but allows for a simpler hardware architecture.

The fact that for the reconstruction the model for the sampling kernel and the propagation are separable allows to easily tailor the imaging pipeline to the requirements of different scenarios and independently optimize the respective implementations. Matrix-free implementations represent a practical approach for this even if the employed algorithm requires additional quantities of the underlying matrix, as illustrated for the approximation of the largest singular value necessary in \gls{fista}. In terms of modeling, it is beneficial to use the analytic signal instead of the widely used real-valued model of the RF signal originating from~\cite{demirli_ieeeuffc_2001}. The increasing demands in computation complexity are compensated by the improved imaging quality by actually approximating a solution to the inverse problem compared to only calculating an image based on a simple heuristic.

The numerical simulations are in agreement with the theoretical observations. This demonstrates that the question of which and how many Fourier coefficients are needed can be directly answered by evaluating the \gls{crb}, which greatly simplifies the parameter specification for a given target scenario. To conclude, the presented results indicate that the use of \gls{cs} is beneficial in the context of synthetic aperture ultrasound \gls{ndt}. In addition, the proposed sampling strategies can also be straightforwardly included into a multi-channel setup and combined with spatial sub-sampling (cf.~\cite{perez_ieeeius_2019}).
%
%
\IEEEpeerreviewmaketitle

\bibliographystyle{IEEEtran}
\bibliography{references.bib}
%

%

\begin{IEEEbiography}%
        [{\includegraphics[width=1in,height=1.25in,
        trim={{1.2cm} {0} {0} {0}},
        clip,keepaspectratio]{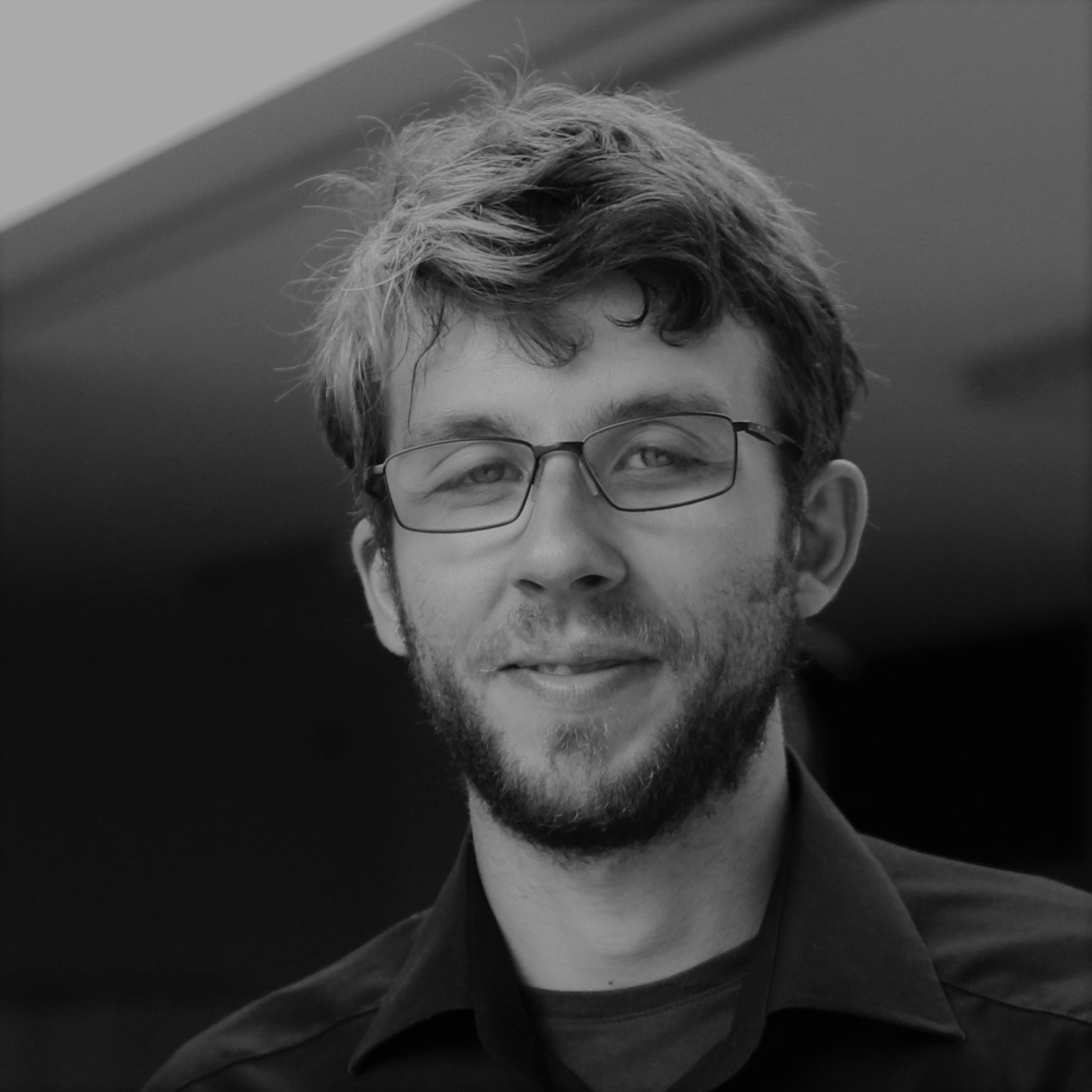}}]%
        {Jan~Kirchhof}
(S’16) received the M.Sc.~degree in media technology from Technische Universität Ilmenau, Ilmenau, Germany, in 2015, where he is currently pursuing the doctorate degree. From 2015 to 2018, he has been with the Electronic~Measurements~and~Signal~Processing~(EMS)~group at Technische~Universität~Ilmenau. Since 2018, he has been with the SigMaSense group at Fraunhofer Institute for Nondestructive Testing IZFP. His current research interest include array signal processing, compressed sensing, and sparse signal recovery for the application in nondestructive testing with ultrasound.
\end{IEEEbiography}

\begin{IEEEbiography}%
  [{\includegraphics[width=1in,height=1.25in,
  trim={{0} {0} {0} {0}},
  clip,keepaspectratio]{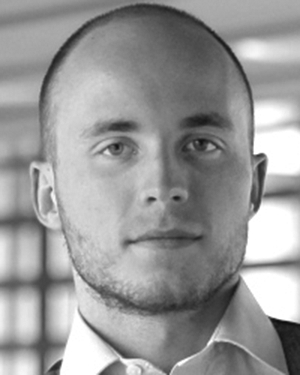}}]%
        {Sebastian~Semper}
studied mathematics at Technische Universität Ilmenau, (TU Ilmenau), Ilmenau, Germany. He received the Master of Science degree in 2015. Since 2015, he has been a Research Assistant with the Electronic Measurements and Signal Processing Group, which is a joint research activity between the Fraunhofer Institute for Integrated Circuits IIS and TU Ilmenau, Ilmenau, Germany, where he is currently working toward the Ph.D. in the field of signal processing. His research focuses on high resolution parameter estimation for channel sounding and radar, compressed sensing in ultrasound and general purpose software for signal processing.
\end{IEEEbiography}

\begin{IEEEbiography}%
        [{\includegraphics[width=1in,height=1.25in,clip,keepaspectratio]{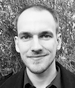}}]%
        {Christoph~W.~Wagner}
        received the M.Sc.~degree in electrical~engineering~and~information~technology from Technische~Universität~Ilmenau, Ilmenau, Germany, in~2016, where he is currently pursuing the doctorate degree. Since~2016, he has been with the Electronic~Measurements~and~Signal~Processing~(EMS)~group at Technische~Universität~Ilmenau. His current research interests include the integrated circuit realization of sub-Nyquist signal acquisition schemes up to mm-Wave frequencies and investigating the properties of alternative analogue-mixed-signal acquisition architectures in integrated CMOS and Bi-CMOS technology for RADAR, GNSS and array signal processing applications.
\end{IEEEbiography}

\begin{IEEEbiography}%
		[{\includegraphics[width=1in,height=1.25in,clip,keepaspectratio]{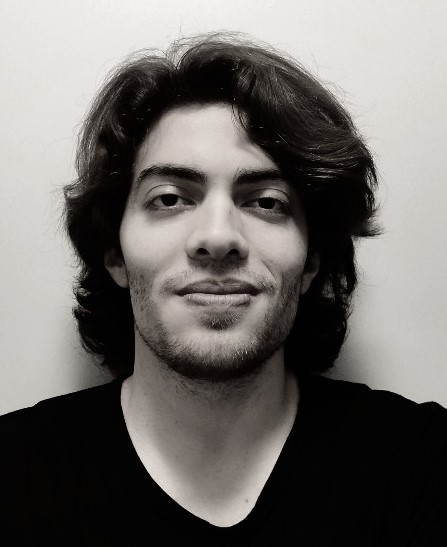}}]%
        {Eduardo~Pérez}
(S’19) received the M.Sc.~degree in communications~and~signal~processing at the Technische Universität Ilmenau, Ilmenau, Germany, in 2019, where he currently pursues a Ph.D. in signal~processing. He joined the SigMaSense group at the Fraunhofer Institute for Nondestructive Testing IZFP and the Electronic~Measurements~and~Signal~Processing~(EMS)~group at Technische~Universität~Ilmenau in 2020. His research interests include array signal processing, compressed sensing and machine learning for ultrasound nondestructive testing.
\end{IEEEbiography}

\begin{IEEEbiography}%
  [{\includegraphics[width=1in,height=1.25in,clip,keepaspectratio]{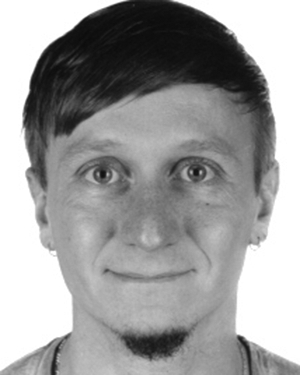}}]%
        {Florian~Römer}
(S’04–M’13–SM’16) studied computer engineering at the Technische Universität Ilmenau (TU Ilmenau), Ilmenau, Germany, and McMaster University, Hamilton, ON, Canada. He received the Diplom-Ingenieur degree in communications engineering and the Doctoral (Dr.-Ing.) degree in electrical engineering from the TU Ilmenau in October 2006 and October 2012, respectively. From December 2006 to September 2012, he was a Research Assistant with the Communications Research Laboratory, TU Ilmenau. In October 2012, he joined the Digital Broadcasting Research Group, a joint research activity between the Fraunhofer Institute for Integrated Circuits and TU Ilmenau, as a Postdoctoral Research Fellow. In January 2018, he joined the Fraunhofer Institute for Nondestructive Testing, where he is currently leading the SigMaSense Group with a research focus on innovative sensing and signal processing for material diagnostics and nondestructive testing. He is the recipient of the Siemens Communications Academic Award for his diploma thesis in 2006 and the EURASIP best dissertation award in 2016 for his dissertation. He was a member of the organizing committee of the 19th International Workshop on Smart Antennas 2015, in Ilmenau, Germany, as well as the IEEE Statistical Signal Processing Workshop 2018, in Freiburg, Germany.
\end{IEEEbiography}

\begin{IEEEbiography}%
  [{\includegraphics[width=1in,height=1.25in,clip,keepaspectratio]{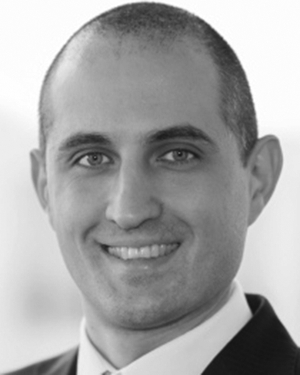}}]
        {Giovanni~Del~Galdo}
(M’12) received the Laurea degree in telecommunications engineering from the Politecnico di Milano, Milan, Italy, in 2002, and the Doctoral degree in MIMO channel modeling for mobile communications from Technische Universität Ilmenau (TU Ilmenau), Ilmenau, Germany, in 2007. He then joined the Fraunhofer Institute for Integrated Circuits, Erlangen, Germany, focusing on audio watermarking and parametric representations of spatial sound. Since 2012, he has been leading a joint research group composed of a Department at Fraunhofer Institute for Integrated Circuits IIS and, as a Full Professor, a Chair with TU Ilmenau on the research area of electronic measurements and signal processing. His current research interests include the analysis, modeling, and manipulation of multidimensional signals, over-the-air testing for terrestrial and satellite communication systems, and sparsity promoting reconstruction methods.
\end{IEEEbiography}






\end{document}